\documentclass[11pt, a4paper, twoside]{article}
\usepackage{graphicx, amsfonts, amsmath, enumerate, tabularx, latexsym, hyperref, multirow, float, cite, caption}
\usepackage{booktabs}
\usepackage[utf8]{inputenc}
\usepackage[T1]{fontenc}
\usepackage[margin=2.5cm, bindingoffset=0.5cm]{geometry}

\newtheorem{definition}{Definition}
\newcommand{\sign}{\mathop{\mathrm{sign}}}

\title{Bifurcations in Ratra-Peebles quintessence models and~their~extensions}
\author{Franciszek Humieja \\
\footnotesize{Astronomical Observatory, Jagiellonian University, Orla 171, 30-244 Kraków, Poland} \\ \\
Marek Szydłowski \\
\footnotesize{Astronomical Observatory, Jagiellonian University, Orla 171, 30-244 Kraków, Poland} \\
\footnotesize{Mark Kac Complex Systems Research Centre, Jagiellonian University, {\L}ojasiewicza 11, 30-348 Kraków, Poland}
}

\date{}
   
\begin{document}
\maketitle

\begin{abstract}
We have used the dynamical system approach in order to investigate the dynamics of cosmological models of the flat Universe with a non-minimally coupled canonical and phantom scalar field and the Ratra-Peebles potential. Applying methods of the bifurcation theory we have found three cases for which the Universe undergoes a generic evolution emerging from either the de~Sitter or the static Universe state and finishing at the de~Sitter state, without the presence of the initial singularity. This generic class of solutions explains both the inflation and the late-time acceleration of the Universe. In this class inflation is an endogenous effect of dynamics itself.
\end{abstract}

\section{Introduction}
The main aim of cosmology is to study the structure and the evolution of the Universe at the large scale. In this context a crucial role plays the notion of a cosmological model obtained from general relativity. If we assume a spacetime with the maximal symmetry of the space-like section of constant time then the evolution of the Universe can be analysed with tools of dynamical systems methods. In principle, there are two advantages of such a treatment of the dynamics. Firstly, evolutional paths are represented in a geometrical way in the phase space---the space of all states of the system under consideration. Secondly, it gives many possibilities of visualisation how solutions (represented in the phase space by trajectories) depend on the initial conditions. Are they exceptional or typical in the phase space? This question is strictly related to a problem of the stability of solutions.

Right-hand sides of the dynamical system depend not only on the state variables (which form the vector of state) but also on parameters. These theoretical parameters are present in any effective theory of the Universe. We believe that their interpretation will emerge if we find more fundamental theory giving us better insight into their nature. Of course, theoretical investigations should be complemented with a statistical estimation of model parameters from the astronomical and astrophysical data.

From the theoretical point of view, it is important to know how the structure of the phase space changes under variation of model parameters. For such an investigation the bifurcation theory is dedicated. Its methods allows to obtain some critical (bifurcation) values of model parameters for which a topological structure of the phase space (modulo homeomorphism preserving the direction of time along trajectories) is not equivalent in the topological sense. If the dynamical system, decribing the evolution of the physical system, undergoes the bifurcation at some value of its parameter, one can classify all possible evolutional paths for all theoretically admissible values of parameters. Hence, in principle, two qualitatively different types can be distinguished. The first case is a the model with the parameter at the bifurcation value which decribes some state, epoch, of the Universe evolution. In this sense the model is fine-tuned as some exact value of the model parameter is required to describe the Universe dynamics. The second case is a model with other, non-bifurcation values of its parameter. The dynamics of the model is qualitatively indifferent with respect to the choice of the value of the model parameter.

Dynamical systems methods are a powerful tool to analyse the dynamics of physical systems and the application of these methods to analyse the cosmological models is very important as the stability conditions allow to constrain or even rule out some models \cite{Bahamonde:2017ize}. The dynamical analysis can be enriched further with using the biffurcation theory methods. Recently, the problem of bifurcations in the FRW cosmology with perfect fluid and the cosmological constant was investigated in the work of Kohli and Haslam\cite{kohli}. They demonstrated that as the cosmological constant parameter is varied, expanding de Sitter and contracting de Sitter universes emerge via fold bifurcation. This type of bifurcation occurs in a neighbourhood of the Minkowski spacetime in the phase space. Also, bifurcations play important role in the idea of the Jungle Universe. Following this approach to study the dynamics of FRW models filled with different barotropic fluids, the jungle cosmological dynamical system assumes the form of a generalised Lotka-Volterra system where the competitive species are replaced with fluids \cite{perez}. The problem of bifurcation is also strictly related to the problem of structural stability which denotes changes of the phase space structure under small change of right-hand sides of the system \cite{Kokarev:2008ba}. It is also interesting to discover the phenomenon of period-doubling bifurcation in strongly anisotropic Bianchi I quantum cosmology\cite{physrev}.

Basic description of the current acceleration of the Universe is provided by the cosmological constant $\Lambda$, whose natural interpretation is the energy of the quantum vacuum. However, this energy could not be constant during the evolution of the Universe from early epochs to the current state. Hence, in order to include both inflation and late time acceleration in one model, it is necessary to introduce for example a tool making the vacuum energy dependent on time. Since the Einstein-Hilbert action must be covariant, one cannot just put $\Lambda(t)$. The solution is to lead in a scalar field $\phi$ together with potential $V(\phi)$, driving the evolution of $\phi$. Such a postulate has been suggested for the first time in \cite{peebles88}, assuming the power-law form of potential, $V(\phi)\propto\phi^n$, where $n=const$. The scalar field approach for describing the late time acceleration of the Universe is known under a name of quintessence \cite{Tsujikawa:2013fta} (for the alternative approach to quintessence see \cite{Kamenshchik:2001cp}).

Apart from standard general relativity terms, it is considered the additional term $\xi\phi^2 R$ in the Einstein-Hilbert action, describing the coupling between the gravity (represented by the Ricci scalar $R$) and the scalar field $\phi$, where $\xi$ is the coupling constant. For $\xi=0$ this coupling is minimal, while for $\xi\neq0$ it is non-minimal. The non-minimal coupling arises from quantum corrections to the scalar field theory \cite{birrel80,allen83,ishikawa83} and is necessary for the renormalisation procedure of the scalar field in the curved space \cite{callan70,freedman-weinberg74}. Moreover, a non-minimal coupling of the scalar field to the gravitation is widely used for modelling inflation \cite{faraoni00,nojiri17}. The generic cosmological evolution in the scalar field cosmology with non-minimal coupling, which does not possess a singularity was investigated by Hrycyna \cite{Hrycyna:2017oug}. These types of evolution correspond to trajectories starting from an unstable de~Sitter state and going to a stable de~Sitter state.

Observational constraints on scalar field cosmological models have been recently investigated in \cite{avsajanishvili18}. Using the Bayesian methods different models of the scalar field cosmology were analysed. The evidence for which as approximation the information criterion BIC was used, indicated that the Ratra-Peebles potential were favoured.

We present the analysis of the dynamical system, describing the evolution of the universe\footnote{The word `universe', started with a lower case letter, means the mathematical model of the Universe.} with the dark energy as a scalar field with a potential. The application of bifurcation theory methods enables us to fully understand the dynamics of the system, which changes its qualitative properties under the variation of its parameters. The dynamical system under consideration includes both the canonical and the phantom scalar field models, which are distinguished by a value of the discrete parameter $\varepsilon$. We assume the power-law form of the potential with the exponent $n$ as a model parameter and the constant non-minimal coupling between the gravity and the scalar field, described by the non-zero parameter~$\xi$.

For the purpose of our investigation, we use elements of local bifurcation theory, whose basics we depict in Section \ref{app}. We discuss local bifurcations of codimension 1, as they appear in models under consideration. For getting more information about the bifurcation theory we refer readers to \cite{Bosi:201938,dercole2011,kuznetsov98,perko91,seydel2009}.

In Section \ref{app} some short introduction to bifurcation theory is given. In Section \ref{dynamical_equations} we derive the dynamical system from cosmological equations and the potential function, in terms of dimensionless phase space variables. In Section \ref{critical_points_non_singular} we find equilibria of the dynamical system, analyse their stability properties, designate conditions they represent the de~Sitter universe, find, on this basis, possible scenarios of the de~Sitter--de~Sitter evolution---which avoids the initial singularity and explains the inflation, as well as the late time acceleration---then prepare the full analysis of bifurcations of the local stability of equilibria for these scenarios. Next, Section \ref{phase_portraits} is devoted to the preparation the phase portraits of the system for non-singular scenarios on the Poincar\'e sphere. Finally, in Section \ref{evolution_of_physical}, we plot the evolution of some physical parameters over the time, within non-singular evolutionary scenarios.

\section{Elements of applied bifurcation theory}
\label{app}

Let us start with some basics on the applied bifurcation theory. First, we give necessary, from the point of bifurcation theory's view, notions of dynamical system theory. Then, we discuss definitions related with the topological equivalence of dynamical systems and bifurcations. Finally, we present a basic description of most important types of local bifurcations. We limit the discussion to bifurcations of codimension one in two-dimensional dynamical systems since this is the only kind of bifurcations that appears in our analysis. Definitions \ref{app_def_dynamical_system}--\ref{app_def_codimension} are quoted from Kuznetsov's book \cite{kuznetsov98}.

\subsection{Dynamical systems}
\label{app_dynamical_systems}
In \emph{deterministic processes}, future and past states of a system can be revealed by knowing its present state and a law describing the evolution. The mathematical formalisation of this fact is the notion of a \emph{dynamical system}.

A set $X$ of all possible \emph{states} of a dynamical system is called a \emph{state space} (or \emph{phase space}). Usually, one distinguishes finite-dimensional systems defined in $X=\mathbb R^n$ from those defined on manifolds. The state $x_t\in X$ of a system changes with time $t\in T$, where $T$ is a number set. In case $T=\mathbb R^1$ a system is called a \emph{continuous-time dynamical system}, while for $T=\mathbb Z$ it is a \emph{discrete-time dynamical system}.

An evolution law of a dynamical system determines the state $x_t$ of the system at time $t$, provided the \emph{initial state $x_0$} is known. Description of the evolution is given by an \emph{evolution operator $\phi^t$}, which is a map
$$\phi^t \colon X\longmapsto X,$$
transforming an initial state $x_0\in X$ into some state $x_t\in X$ at time $t$,
$$x_t=\phi^tx_0.$$

Deterministic nature of dynamical systems is reflected in following two properties of the evolution operator. First of all,
\begin{equation}
\label{evo_op_prop1}
\phi^0=\mathrm{id},
\end{equation}
where $\mathrm{id}$ is the identity map on $X$, $\mathrm{id}\;x=x$ for all $x\in X$. The second property is
\begin{equation}
\label{evo_op_prop2}
\phi^{t+s}=\phi^t\circ\phi^s,
\end{equation}
which implies $x_{t+s}=\phi^t\!\left(\phi^sx_0\right)$.

Now we are ready to give the formal definition of the dynamical system, which comes from \cite{kuznetsov98}.
\begin{definition}
\label{app_def_dynamical_system}
A \emph{dynamical system} is a triple $\{T,X,\phi^t\}$, where $T$ is a time set, $X$ is a state space, and $\phi^t:X\mapsto X$ is a family of evolution operators parametrised by $t\in T$ and satisfying properties (\ref{evo_op_prop1}) and (\ref{evo_op_prop2}).
\end{definition}

In case of continuous-time systems, a family $\left\{\phi^t\right\}_{t\in T}$ of evolution operators is called a \emph{flow}. Moreover, in case of both continuous- and discrete-time systems, an \emph{orbit $\Gamma$} is an ordered subset of the state space $X$,
$$\Gamma(x_0)=\left\{x\in X \colon x=\phi^tx_0,\mbox{ for all }t\in T\mbox{ such that }\phi^tx_0\mbox{ is defined}\right\},$$
Orbits in the state space of a dynamical system compose a \emph{phase portrait}, which is the geometrical representation of the dynamical system.

Very often one identifies a continuous-time dynamical system with \emph{differential equations} describing implicitly a law of evolution. Suppose that the state space of a system is $X=\mathbb R^n$. Then, differential equations related to the system are
\begin{equation}
\label{diff_eq_param}
\dot x=f(x,\alpha),
\end{equation}
where $\dot x=\mathrm dx/\mathrm dt$, the state $x=(x_1,\dots,x_n)\in\mathbb R^n$, $\alpha=(\alpha_1,\dots,\alpha_m)\in\mathbb R^m$ is the parameter vector, and $f:\mathbb R^n\times\mathbb R^m\ni(x,\alpha)\mapsto(f_1,\dots,f_n)\in\mathbb R^n$ is a function (a vector field) of the $C^k$~class, with $k\geq1$. We will frequently refer to dynamical systems without underlining its dependence on the parameter vector $\alpha$. Thus, differential equations will have the form
\begin{equation}
\label{diff_eq}
\dot x=f(x).
\end{equation}

Let us quote the following three important definitions, taken from \cite{kuznetsov98}, which are related to dynamical systems. First of them is an equilibrium, which is the special case of an orbit.
\begin{definition}
A point $x^0\in X$ is called an \emph{equilibrium} (or a \emph{fixed point}) if $\phi^tx^0=x^0$ for all $t\in T$.
\end{definition}

An equilibrium is thus a point in which the system remains forever. For differential equations (\ref{diff_eq}) in an equilibrium $x^0$ we have
$$f(x^0)=0.$$

Subsequent definitions concern the notion of another special kind of orbits---cycles.
\begin{definition}
A \emph{cycle} is a periodic orbit, namely a non-equilibrium orbit $L_0$, such that each point $x_0\in L_0$ satisfies $\phi^{t+T_0}x_0=\phi^tx_0$ with some $T_0>0$, for all $t\in T$.
\end{definition}

\begin{definition}
A cycle of a continuous-time dynamical system, in a neighbourhood of which there are no other cycles, is called a \emph{limit cycle}.
\end{definition}

\subsection{Topological equivalence and bifurcations}
\label{app_topological_equivalence}
Before introducing the notion of bifurcation, we should clarify what it means that phase portraits of two dynamical systems have the same qualitative features. For this purpose, let us quote from \cite{kuznetsov98} the definition of topological equivalence of two dynamical systems.
\begin{definition}
\label{topo_equiv}
A dynamical system $\{T,\mathbb R^n,\phi^t\}$ is called \emph{topologically equivalent} to a dynamical system $\{T,\mathbb R^n,\psi^t\}$ if there is a homeomorphism $h:\mathbb R^n\mapsto\mathbb R^n$ mapping orbits of the first system onto orbits of the second system, preserving the direction of time.
\end{definition}

Dynamical systems are very often studied in a vicinity of an equilibrium since it is possible to lead useful linear approximations there. Therefore, one may elaborate the topological classification of phase portraits near equilibrium points and hence the need for the following modification of Definition \ref{topo_equiv}.
\begin{definition}
A dynamical system $\{T,\mathbb R^n,\phi^t\}$ is called \emph{locally topologically equivalent} near an equilibrium $x_0$ to a dynamical system $\{T,\mathbb R^n,\psi^t\}$ near an equilibrium $y_0$ if there exists a homeomorphism $h:\mathbb R^n\mapsto\mathbb R^n$ that is
\begin{enumerate}[(i)]
\item defined in a small neighbourhood $U\subset\mathbb R^n$ of $x_0$;
\item satisfies $y_0=h(x_0)$;
\item maps orbits of the first system in $U$ onto orbits of the second system in $V=h(U)\subset\mathbb R^n$, preserving the direction of time.
\end{enumerate}
\end{definition}

Consider the dynamical system (\ref{diff_eq_param}) dependent on parameters $\alpha\in\mathbb R^m$. As parameters vary, the phase portrait of the system also varies. This results in two possibilities: either the system remains topologically equivalent to the original one, or its topology changes.
\begin{definition}
The appearance of topologically nonequivalent phase portrait under variation of parameters is called a \emph{bifurcation}.
\end{definition}

Thus, a bifurcation is a change of the topological type of the system. A value of parameters at which this change happens is called a \emph{bifurcation (critical) value}.

If a bifurcation occurs when we fix {\sl any} small neighbourhood of an equilibrium, then it is called a \emph{local bifurcation} (or a \emph{bifurcation of an equilibrium}). Otherwise, when a bifurcation cannot be determined by analysing only a small vicinity of an equilibrium, it is a \emph{global bifurcation}.

It is often convenient to visualise how the occurrence of bifurcations depends on parameters of the system. For this purpose, let us introduce the definition of a bifurcation diagram taken from \cite{kuznetsov98}.
\begin{definition}
A \emph{bifurcation diagram} of the dynamical system is a stratification of its parameter space induced by the topological equivalence, together with representative phase portraits for each stratum.
\end{definition}

In the simplest case, the bifurcation diagram consists of a finite number of regions in the parameter space $\mathbb R^m$, inside which the phase portrait is topologically equivalent. These regions are separated by \emph{bifurcation boundaries}, which are smooth submanifolds in $\mathbb R^m$ (e.g. curves, surfaces).

\begin{definition}
\label{app_def_codimension}
A \emph{codimension} of a bifurcation is the difference between the dimension of the parameter space and the dimension of the corresponding bifurcation boundary.
\end{definition}

The equivalent and more practical definition says that a codimension is the number of independent conditions determining the bifurcation.

Bifurcations are divided into many types that differ in their course. It is possible to lead the most simple differential equations that represent a given type of bifurcation. These equations are called \emph{normal forms}. In case of local bifurcations, every dynamical system having a local bifurcation of a specified type is locally topologically equivalent near an equilibrium to the normal form of this type of bifurcation.

\subsection{Local bifurcations of codimension one in continuous-time systems}
\label{app_local_bifurcations}
Let us introduce most important types of local bifurcations in continuous-time systems of at most two variables. We limit the discussion to bifurcations of codimension one. We present four types of bifurcations. Three of them are bifurcations of the \emph{saddle-node family}: an \emph{elementary saddle-node}, a \emph{transcritical}, and a \emph{pitchfork} bifurcation, while the last is a \emph{Hopf bifurcation}.

\begin{figure}[tb]
	\captionsetup{font=small}
	\small
	\centering
	\includegraphics[width=0.45\textwidth]{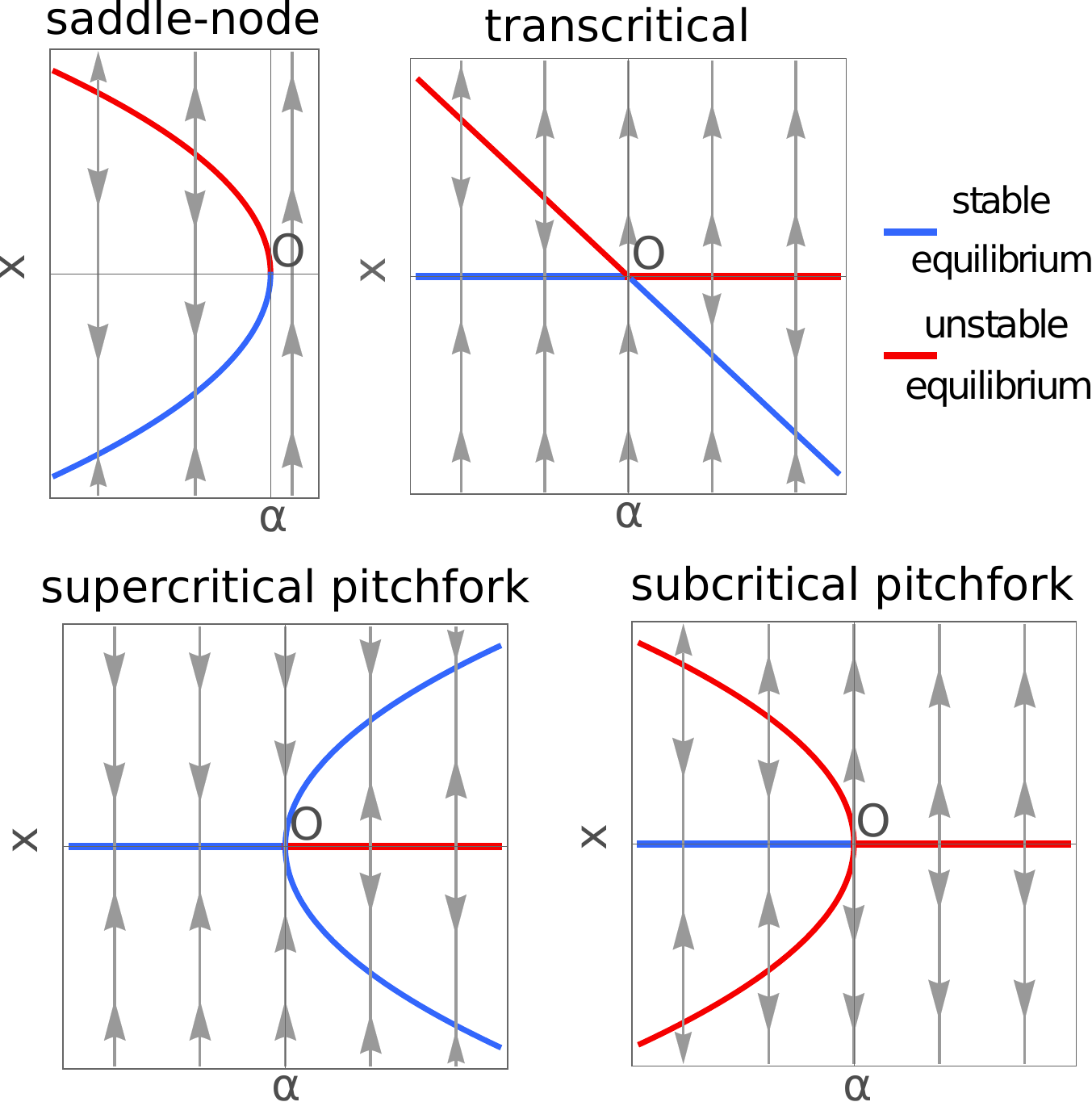}
	\caption{Bifurcation diagrams corresponding to the normal forms (\ref{normal_form_esn}--\ref{normal_form_sp}). Bifurcations take place in the origin $O$.}
	\label{bifurcation-types}
\end{figure}

Bifurcations of the \emph{saddle-node family} (also called \emph{fold bifurcations}) arise as a collision of equilibria, when a real eigenvalue crosses zero in a response to a change of a parameter. In the case of an \emph{elementary saddle-node bifurcation}, two equilibria---one stable, the other unstable---coalesce and then disappear. In the case of a \emph{transcritical bifurcation}, two equilibria---one stable, the other unstable---coalesce and then separate again, exchanging their stability properties. Finally, in the case of a \emph{pitchfork bifurcation} there are two possibilities: in first of them, called \emph{supercritical}, three equilibria---one unstable, surrounded by two stable---coalesce into one stable equilibrium, while in the other, called \emph{subcritical}, three equilibria---one stable, surrounded by two unstable---coalesce into one unstable equilibrium. Normal forms for saddle-node family of bifurcations are 
\begin{eqnarray}
\label{normal_form_esn}
\mbox{elementary saddle-node,} & \quad & \dot x=\alpha\pm x^2, \\
\mbox{transcritical,} & \quad & \dot x=\alpha x\pm x^2, \\
\mbox{supercritical pitchfork,} & \quad & \dot x=\alpha x-x^3, \\
\label{normal_form_sp}
\mbox{subcritical pitchfork,} & \quad & \dot x=\alpha x+x^3,
\end{eqnarray}
where $\alpha$ is the parameter with the critical value equal to zero. Bifurcation diagrams of the foregoing normal forms are shown in Figure \ref{bifurcation-types}.

A \emph{Hopf bifurcation} (known also as an \emph{Andronov-Hopf bifurcation}) arises, when real parts of two complex and conjugate eigenvalues crosses zero. It presents a collision of a stable or unstable equilibrium with a cycle, which shrinks to a point when the collision occurs. The normal form for this kind of bifurcation is
\begin{equation}
\begin{aligned}
\dot x_1&=\alpha x_1-x_2-x_1(x_1^2+x_2^2),\\
\dot x_2&=x_1+\alpha x_2-x_2(x_1^2+x_2^2),
\end{aligned}
\end{equation}
which in polar coordinates $(\rho,\varphi)$ is
\begin{equation}
\begin{aligned}
\dot\rho&=\rho(\alpha-\rho^2),\\
\dot\varphi&=1,
\end{aligned}
\end{equation}
where $\alpha$ is the parameter with the critical value equal to zero.

\section{Dynamical equations}
\label{dynamical_equations}

The scalar field approach describes the dark energy by the {\it scalar field $\phi$}, which is affected by the {\it potential $U(\phi)>0$}. The total action is
\begin{equation}
	\label{total_action}
	S=S_g+S_\phi,
\end{equation}
where
\begin{equation}
	S_g=\frac{1}{2\kappa^2}\int\mathrm d^4x\sqrt{-g}R
\end{equation}
is the Einstein-Hilbert action, and
\begin{equation}
	S_\phi=-\frac{1}{2}\int\mathrm d^4x\sqrt{-g}\left[\varepsilon\nabla^\alpha\phi\nabla_\alpha\phi+\varepsilon\xi R\phi^2+2U(\phi)\right],
\end{equation}
with $\varepsilon=\pm1$ corresponding to the {\it canonical} and the {\it phantom scalar field} respectively, $\kappa^2=8\pi G$ and $c=\hbar=1$. The parameter $\xi$ is called the {\it coupling parameter}; if $\xi\neq0$, the scalar field is {\it non-minimally coupled to the gravity}. Otherwise, the scalar field is {\it minimally coupled to the gravity}.

The variation of the total action (\ref{total_action}) with respect to the metric tensor $g^{\mu\nu}$ yields the field equation
\begin{equation}
	R_{\mu\nu}-\frac{1}{2}g_{\mu\nu}R=\kappa^2T_{\mu\nu}^{(\phi)},
\end{equation}
where the energy-momentum tensor for the non-minimally coupled scalar field is given by
\begin{equation}
	\begin{aligned}
		T_{\mu\nu}^{(\phi)}=&\varepsilon\nabla_\mu\phi\nabla_\nu\phi-\frac{1}{2}\varepsilon g_{\mu\nu}\nabla^\alpha\phi\nabla_\alpha\phi-g_{\mu\nu}\,U(\phi)+\\
		&+\varepsilon\xi\phi^2\left(R_{\mu\nu}-\frac{1}{2}g_{\mu\nu}R\right)+\varepsilon\xi\left(g_{\mu\nu}\Box\phi^2-\nabla_\mu\nabla_\nu\phi^2\right),
	\end{aligned}
\end{equation}
while the variation with respect to the scalar field $\phi$ gives the Klein-Gordon equation
\begin{equation}
	\Box\phi-\xi R\phi-\varepsilon U_{,\phi}=0,
\end{equation}
where $U_{,\phi}:=\mathrm dU/\mathrm d\phi$.

We consider the spatially flat ($k=0$) Friedmann-Lema\^itre-Robertson-Walker (FLRW) metric
\begin{equation}
	\mathrm ds^2=-\mathrm dt^2+a^2(t)\left(\mathrm dx^2+\mathrm dy^2+\mathrm dz^2\right),
\end{equation}
where $a(t)$ is the scale factor. This produces the Friedmann equation in the form
\begin{equation}
	\label{friedmann}
	\frac{3}{\kappa^2}H^2=\rho_\phi=\frac{1}{2}\varepsilon\dot\phi^2+U+3\varepsilon\xi H^2\phi^2+6\varepsilon\xi H\phi\dot\phi,
\end{equation}
the acceleration equation
\begin{equation}
	\label{acceleration}
	-\frac{1}{\kappa^2}\left(2\dot H+3H^2\right)=p_\phi=\frac{\frac{1}{2}\varepsilon\dot\phi^2(1-4\xi)+6\varepsilon\xi^2\phi^2H^2+2\varepsilon\xi H\phi\dot\phi-U+2\xi\phi U_{,\phi}}{1-\varepsilon\kappa^2\xi\phi^2(1-6\xi)},
\end{equation}
and the Klein-Gordon equation
\begin{equation}
	\label{klein_gordon}
	\ddot\phi+3H\dot\phi+6\xi\phi\left(\dot H+2H^2\right)+\varepsilon U_{,\phi}=0,
\end{equation}
where $\rho_\phi$ is the energy density and $p_\phi$ is the pressure of the scalar field, which are connected by the linear barotropic equation of state
\begin{equation}
	p_\phi=w_\phi\,\rho_\phi,
\end{equation}
where $w_\phi$ is the equation of state parameter.

Let us note, that from the Friedmann equation (\ref{friedmann}) we see that the total energy density of the scalar field consists of the kinetic energy density term $\frac{1}{2}\varepsilon\dot\phi^2$, the potential energy density term $U$, and the term related to the gravity-scalar field interaction $3\varepsilon\xi H\phi(H\phi+2\dot\phi)$. Dividing this equation by the critical density $3H^2\kappa^{-2}$ we obtain an energy conservation equation in the form
\begin{equation}
	\label{energy_conservation}
	1=\Omega_{\phi,\text{kin}}+\Omega_{\phi,\text{pot}}+\varepsilon\kappa^2\xi\phi\left(\phi+\frac{2\dot\phi}{H}\right),
\end{equation}
where
\begin{equation}
	\label{kinetic_parameter}
	\Omega_{\phi,\text{kin}}:=\frac{\varepsilon\kappa^2\dot\phi^2}{6H^2}
\end{equation}
is the scalar field kinetic energy parameter and
\begin{equation}
	\label{potential_parameter}
	\Omega_{\phi,\text{pot}}:=\frac{\kappa^2U}{3H^2}
\end{equation}
is the scalar field potential energy parameter.

We assume the inverse power-law form of the potential function (known as the {\it Ratra-Peebles potential} \cite{peebles88,ratra87})
\begin{equation}
	U(\phi)=\frac{M^{n+4}}{\phi^n},
\end{equation}
where $n$ is a dimensionless parameter, and $M>0$ is a dimensional constant. The pair of parameter values $\varepsilon=+1$ and $\xi=0$, together with the usage of the power-law form of the potential, refer to the standard {\it Ratra-Peebles quintessence model}, while other values constitute extensions of this model.

Let us introduce following dimensionless real phase space variables
\begin{equation}
	\label{vars}
	u=\frac{\dot\phi}{H\phi},\quad v=\frac{\sqrt 6}{\kappa}\frac{1}{\phi}.
\end{equation}
Applying new variables, we obtain the following condition from the Friedmann equation~(\ref{friedmann}) 
\begin{equation}
	\label{friedmann_uv}
	\frac{M^{n+4}}{H^2\phi^{n+2}}=\frac{1}{2}v^2-\frac{1}{2}\varepsilon u^2-3\varepsilon\xi(1+2u),
\end{equation}
which, in terms of energy parameters (\ref{kinetic_parameter}--\ref{potential_parameter}), reads
\begin{equation}
	\label{energy_conservation_uv}
	1=\Omega_{\phi,\text{kin}}+\Omega_{\phi,\text{pot}}+6\varepsilon\xi\frac{1+2u}{v^2},
\end{equation}
where
\begin{equation}
	\label{energy_parameters}
	\Omega_{\phi,\text{kin}}=\varepsilon\frac{u^2}{v^2}\quad\mbox{and}\quad\Omega_{\phi,\text{pot}}=\frac{\kappa^2M^{n+4}}{3H^2\phi^n}.
\end{equation}
The acceleration equation (\ref{acceleration}) yields
\begin{equation}
	\label{acceleration_uv}
	\frac{\dot H}{H^2}=-\frac{3}{2}(1+w_\phi),
\end{equation}
where the equation of state parameter, in terms of new variables, is
\begin{equation}
	\label{eos}
	w_\phi=\frac{2}{3}\frac{u^2\left[1-\xi(2-n)\right]+4\xi u(2+3\xi n)-\frac{1}{2}\varepsilon v^2(1+2\xi n)+3\xi\left[1+2\xi(1+n)\right]}{\frac{1}{3}\varepsilon v^2-2\xi(1-6\xi)}.
\end{equation}
The left-hand side of equation (\ref{acceleration_uv}) is equal to $-q-1$, where $q=-\ddot aa\dot a^{-2}$ is the deceleration parameter; $q<0$ for accelerated expansion (or decelerated contraction) of the universe, and $q>0$ for decelerated expansion (or accelerated contraction) of the universe. Hence, we obtain
\begin{equation}
	q=\frac{1}{2}(3w_\phi+1),
\end{equation}
which implies that for $w_\phi<-\frac{1}{3}$ expansion of the universe is accelerated (contraction is decelerated), whereas for $w_\phi>-\frac{1}{3}$ expansion is decelerated (contraction is accelerated).

Finally, using equations (\ref{klein_gordon}), (\ref{friedmann_uv}) and (\ref{acceleration_uv}), we obtain the dynamical system, expressed by dimensionless variables (\ref{vars})
\begin{equation}
	\label{system1}
	\begin{aligned}
		u'=&-\frac{1}{2}u^2(2+n)-\frac{3}{2}u(1+4\xi n)+\frac{1}{2}\varepsilon nv^2-3\xi(1+n)+\\
		&+(6\xi+u)\frac{u^2\left[1-\xi(2-n)\right]+4\xi u(2+3\xi n)-\frac{1}{2}\varepsilon v^2(1+2\xi n)+3\xi\left[1+2\xi(1+n)\right]}{\frac{1}{3}\varepsilon v^2-2\xi(1-6\xi)},\\
		v'=&-uv,
	\end{aligned}
\end{equation}
where $f'=\frac{\mathrm df}{\mathrm d\ln a}=H^{-1}\dot f$.

Formulae for $w_\phi$ (\ref{eos}) and for $u'$ in the system (\ref{system1}) diverge to infinity as
\begin{equation}
	\label{limit}
	v^2\rightarrow6\varepsilon\xi(1-6\xi).
\end{equation}
In order to analyse the dynamics at this limit, we can multiply right-hand sides of the system (\ref{system1}) by the non-negative term $\left[\frac{1}{3}\varepsilon v^2-2\xi(1-6\xi)\right]^2$. This operation will produce a dynamical system which will be determined at the limit (\ref{limit}) and will have the same dynamical properties in other points as the previous system\footnote{Such an operation is also related to the reparameterisation of the time. Consider the system $\mathrm d{\bf x}/\mathrm dt={\bf f}({\bf x})$ and multiply its both sides by a function $\xi({\bf x})>0$. As a result of this operation, we obtain the new system $\mathrm d{\bf x}/\mathrm d\tau=\xi({\bf x}){\bf f}({\bf x})$, where $\mathrm d\tau:=\mathrm dt/\xi({\bf x})$, which has the same equilibria and is topologically equivalent to the original one.} Thus, we obtain the system
\begin{equation}
	\label{system}
	\begin{aligned}
		u'=&\left[-\frac{1}{2}u^2(2+n)-\frac{3}{2}u(1+4\xi n)+\frac{1}{2}\varepsilon nv^2-3\xi(1+n)\right]\left[\frac{1}{3}\varepsilon v^2-2\xi(1-6\xi)\right]^2+\\
		&+(6\xi+u)\left[\frac{1}{3}\varepsilon v^2-2\xi(1-6\xi)\right]\times\\
		&\times\left(u^2\left[1-\xi(2-n)\right]+4\xi u(2+3\xi n)-\frac{1}{2}\varepsilon v^2(1+2\xi n)+3\xi\left[1+2\xi(1+n)\right]\right),\\
		v'=&-uv\left[\frac{1}{3}\varepsilon v^2-2\xi(1-6\xi)\right]^2,
	\end{aligned}
\end{equation}
which is determined also at the limit (\ref{limit}).

From initial assumptions we have $U(\phi)>0$ equivalent to $\phi>0$, which implies $v>0$. It also implies that the left-hand side of equation (\ref{friedmann_uv}) is greater than zero. It yields the following condition of physicality on variables $(u,v)$
\begin{equation}
	\label{condition}
	\frac{1}{2}v^2-\frac{1}{2}\varepsilon u^2-3\varepsilon\xi(1+2u)>0.
\end{equation}

\section{Equilibria, non-singular evolutionary scenarios and bifurcation diagrams of local stability}
\label{critical_points_non_singular}

\renewcommand{\arraystretch}{1.2}
\begin{table}[t!]
	\captionsetup{font=small}
	\small
	\centering
	\caption{Finite equilibria of the system (\ref{system}) with existence conditions.}
	\label{critical-points}
	\begin{tabular}{@{}llll@{}}
		\toprule
		point & $u$ & $v$ & existence \\
		\midrule
		$A$ & $-6\xi-\sqrt{6\xi(6\xi-1)}$ & $0$ & \multirow{2}{*}{$\left(\xi\leq0\lor\xi\geq\frac{1}{6}\right)\ \land\ n\in\mathbb R\ \land\ \varepsilon=\pm1$} \\
		\cmidrule{1-3}
		$B$ & $-6\xi+\sqrt{6\xi(6\xi-1)}$ & $0$ & \\
		\midrule
		$C$ & $\frac{\xi(4+n)}{\xi(2-n)-1}$ & $0$ & $\xi\in\mathbb R\ \land\ n\neq2-\frac{1}{\xi}\ \land\ \varepsilon=\pm1$ \\
		\midrule
		\multirow{2}{*}{$D$} & \multirow{2}{*}{$0$} & \multirow{2}{*}{$\frac{\sqrt{6\xi}}{\sqrt\varepsilon}$} & $\xi\geq0\ \land\ n\in\mathbb R\ \land\ \varepsilon=+1$ \\
		 & & & $\xi\leq0\ \land\ n\in\mathbb R\ \land\ \varepsilon=-1$ \\
		\midrule
		\multirow{4}{*}{$E$} & \multirow{4}{*}{$0$} & \multirow{4}{*}{$\frac{\sqrt{6\xi(4+n)}}{\sqrt{\varepsilon n}}$} & $\xi\geq0\ \land\ \left(n\leq-4\lor n>0\right)\ \land\ \varepsilon=+1$ \\
		 & & & $\xi\leq0\ \land\ -4\leq n<0\ \land\ \varepsilon=+1$ \\
		 & & & $\xi\geq0\ \land\ -4\leq n<0\ \land\ \varepsilon=-1$ \\
		 & & & $\xi\leq0\ \land\ \left(n\leq-4\lor n>0\right)\ \land\ \varepsilon=-1$ \\
		 \midrule
		 \multirow{2}{*}{$F$} & \multirow{2}{*}{any} & \multirow{2}{*}{$\frac{\sqrt{6\xi(1-6\xi)}}{\sqrt\varepsilon}$} & $0\leq\xi\leq\frac{1}{6}\ \land\ n\in\mathbb R\ \land\ \varepsilon=+1$ \\
		 & & & $\left(\xi\leq0\lor\xi\geq\frac{1}{6}\right)\ \land\ n\in\mathbb R\ \land\ \varepsilon=-1$ \\
		\bottomrule
	\end{tabular}
\end{table}

\begin{table}[t!]
	\captionsetup{font=small}
	\small
    \centering
	\caption{Eigenvalues, stability type, equation of state parameter $w_\phi$ value and the condition for the de~Sitter universe for equilibria of system (\ref{system}).}
	\label{critical-points-details}
	\begin{tabular}{@{}lllll@{}}
		\toprule
		point & \multicolumn{1}{l}{eigenvalues} & type & $w_\phi$ & de~Sitter universe \\
		\midrule
		\multirow{2}{*}{$A$} & $\lambda_1=\alpha(\xi)\cdot\beta_+(\xi)$ & \multirow{2}{*}{see Fig. \ref{bif-a}} & \multirow{2}{*}{$1-\frac{4}{3}\beta_+(\xi)$} & \multirow{2}{*}{for $\xi=\frac{3}{16}$} \\
		 & $\lambda_2=6\alpha(\xi)+(n-2)\lambda_1$ & & & \\
		\midrule
		\multirow{2}{*}{$B$} & $\lambda_1=\alpha(\xi)\cdot\beta_-(\xi)$ & \multirow{2}{*}{see Fig. \ref{bif-b}} & \multirow{2}{*}{$1-\frac{4}{3}\beta_-(\xi)$} & \multirow{2}{*}{no} \\
		 & $\lambda_2=6\alpha(\xi)+(n-2)\lambda_1$ & & & \\
		\midrule
		\multirow{2}{*}{$C$} & $\lambda_1=\frac{\alpha(\xi)\cdot\xi(n+4)}{1+\xi(n-2)}$ & \multirow{2}{*}{see Fig. \ref{bif-c}} & \multirow{2}{*}{$-\frac{\xi(n^2+9n+2)+3}{3\xi(n-2)+3}$} & for $\xi=0$ \\
		 & $\lambda_2=-\frac{\alpha(\xi)\cdot[6+\xi(n-2)(n+10)]}{2[1+\xi(n-2)]}$ & & & or $n\in\{-4,-2\}$ \\
		\midrule
		\multirow{2}{*}{$D$} & $\lambda_1=-144\xi^4$ & \multirow{2}{*}{saddle} & \multirow{2}{*}{$\frac{1}{3}$} & \multirow{2}{*}{no} \\
		 & $\lambda_2=576\xi^4$ & & & \\
		\midrule
		$E$ & $\lambda_{1,2}=\gamma(\xi,n)\pm\delta(\varepsilon,\xi,n)$ & see Fig. \ref{bif-e}, \ref{bif-ee} & $-1$ & yes \\
		\midrule
		$F$ & $\lambda_{1,2}=0$ & neutral line & indeterminate* & no \\
		\bottomrule
		\multicolumn{5}{l}{where:} \\
		\multicolumn{2}{l}{$\alpha(\xi):=4\xi^2(1-6\xi)^2$} & \multicolumn{3}{l}{$\gamma(\xi,n):=-\frac{24\xi^2(2+3n\xi)^2}{n^2}$} \\
		\multicolumn{2}{l}{$\beta_\pm(\xi):=6\xi\pm\sqrt{6\xi(6\xi-1)}$} & \multicolumn{3}{l}{$\delta(\varepsilon,\xi,n):=-\frac{8\xi^2(2+3n\xi)\sqrt{-3n\varepsilon(2+3n\xi)[\xi(7n+64)-6]}}{\sqrt{n^5}\sqrt\varepsilon}$} \\
		\multicolumn{5}{p{0.8\textwidth}}{\scriptsize *There exist only one-sided limits of $w_\phi$ as $v$ approaches the line $F$. These limits diverge to $\pm\infty$, depending on the values of $u$, $\xi$, $n$ and $\varepsilon$.} \\
		\bottomrule
	\end{tabular}
\end{table}

In this section, we will find equilibria of the dynamical system derived above, inspect stability of these points and conditions for representing the de~Sitter state by them, distinguish, on this basis, evolutionary scenarios without singularity, and---finally---prepare bifurcation diagrams for these scenarios.

\begin{figure}[t!]
	\captionsetup{font=small}
	\small
	\centering
	\includegraphics[width=0.8\textwidth]{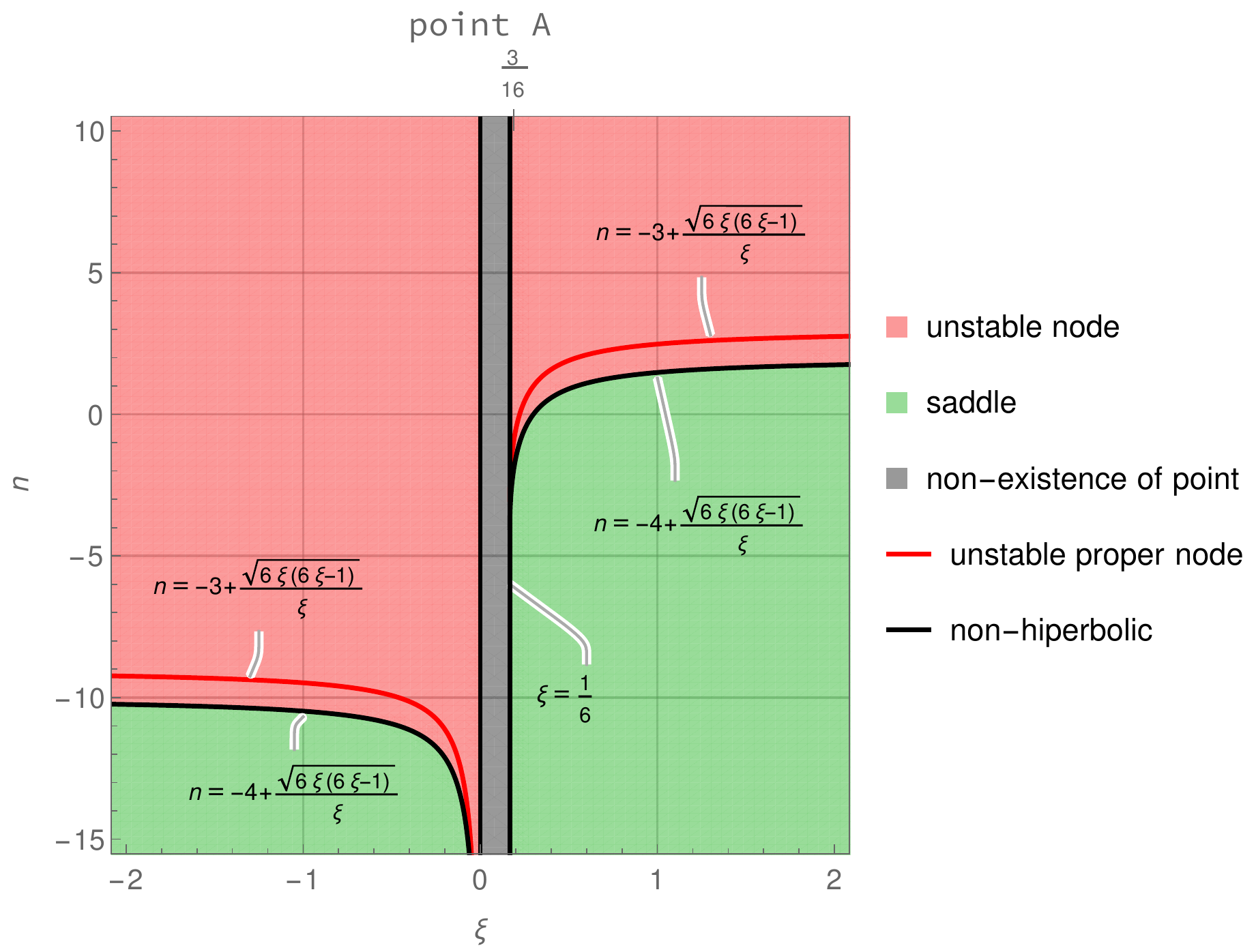}
	\caption{Bifurcation diagram of the local stability of point $A$ for $\varepsilon=\pm1$.}
	\label{bif-a}
	\vspace{30px}
	\includegraphics[width=0.8\textwidth]{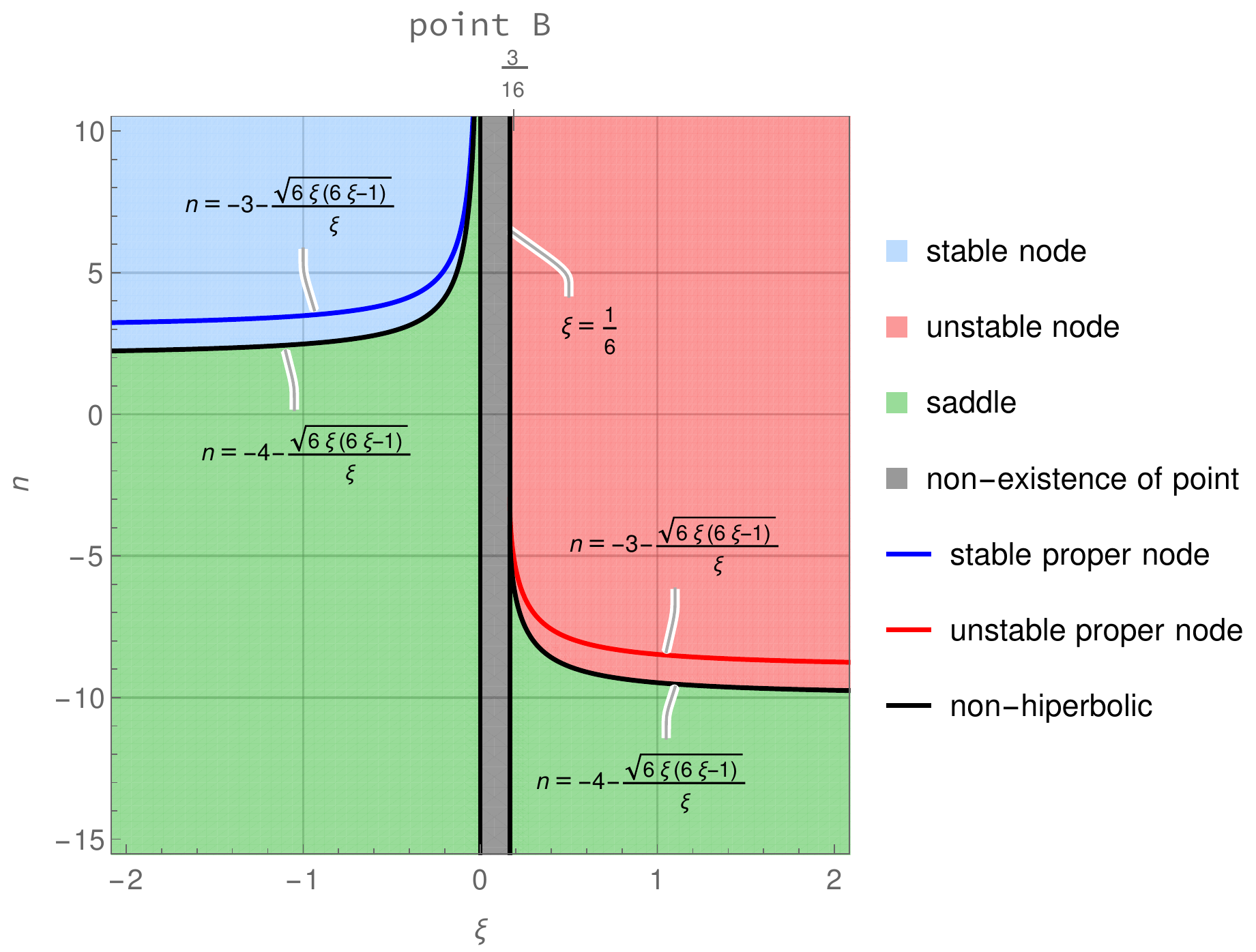}
	\caption{Bifurcation diagram of the local stability of point $B$ for $\varepsilon=\pm1$.}
	\label{bif-b}
\end{figure}

\begin{figure}[t!]
	\captionsetup{font=small}
	\small
	\centering
	\includegraphics[width=0.8\textwidth]{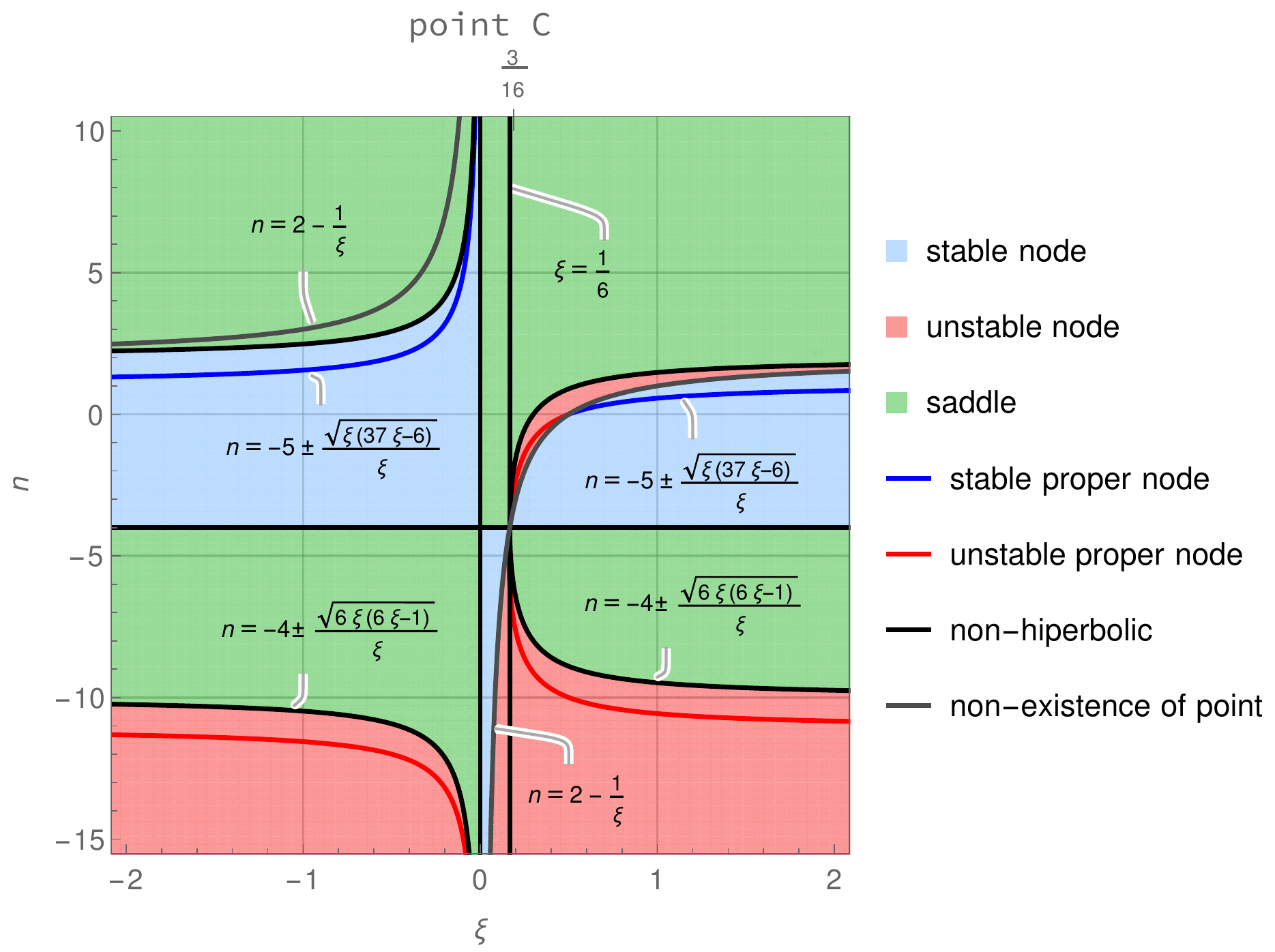}
	\caption{Bifurcation diagram of the local stability of point $C$ for $\varepsilon=\pm1$.}
	\label{bif-c}
	\vspace{30px}
	\includegraphics[width=0.8\textwidth]{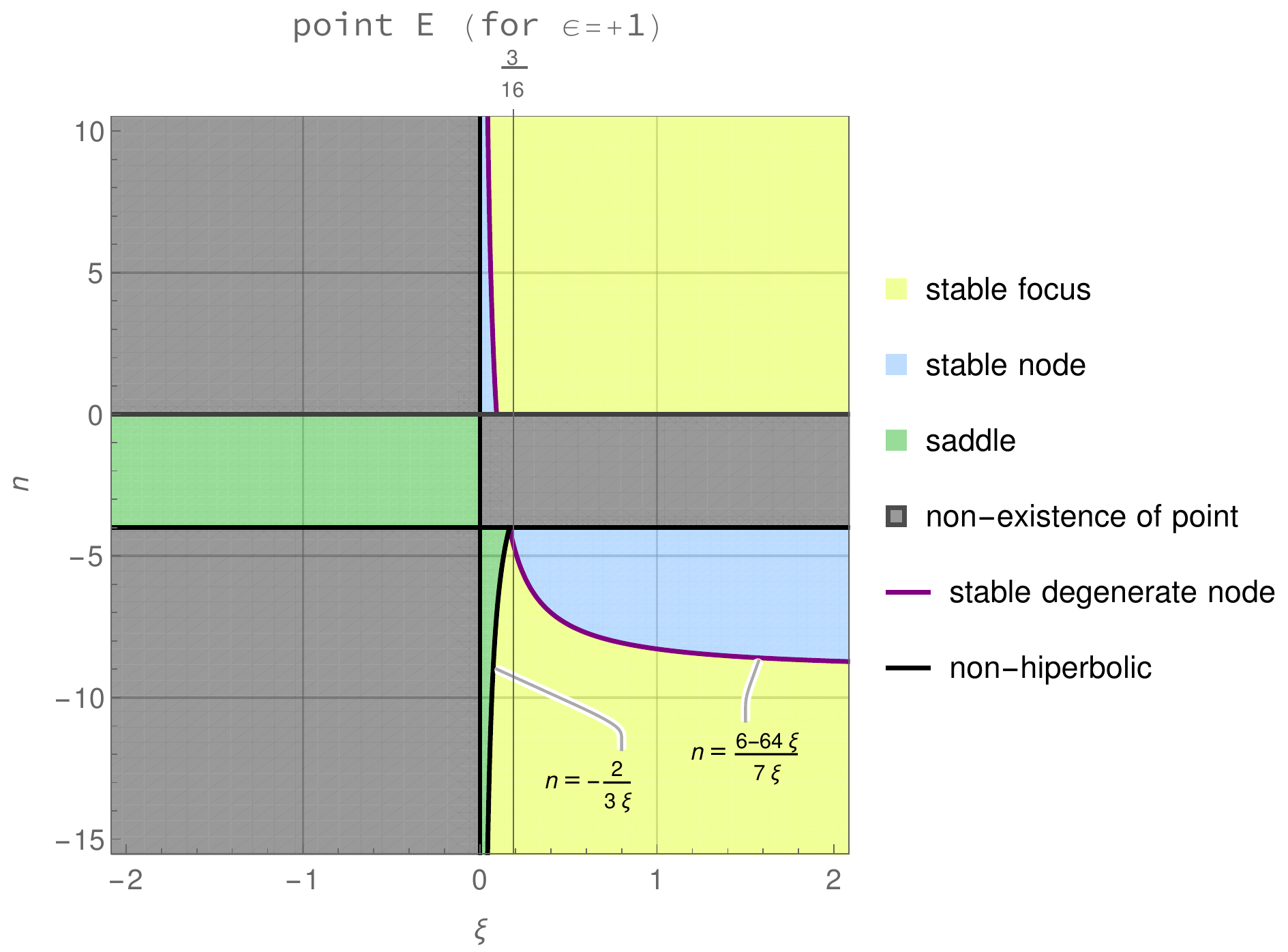}
	\caption{Bifurcation diagram of the local stability of point $E$ for $\varepsilon=+1$.}
	\label{bif-e}
\end{figure}

\begin{figure}[t!]
	\captionsetup{font=small}
	\small
	\centering
	\includegraphics[width=0.8\textwidth]{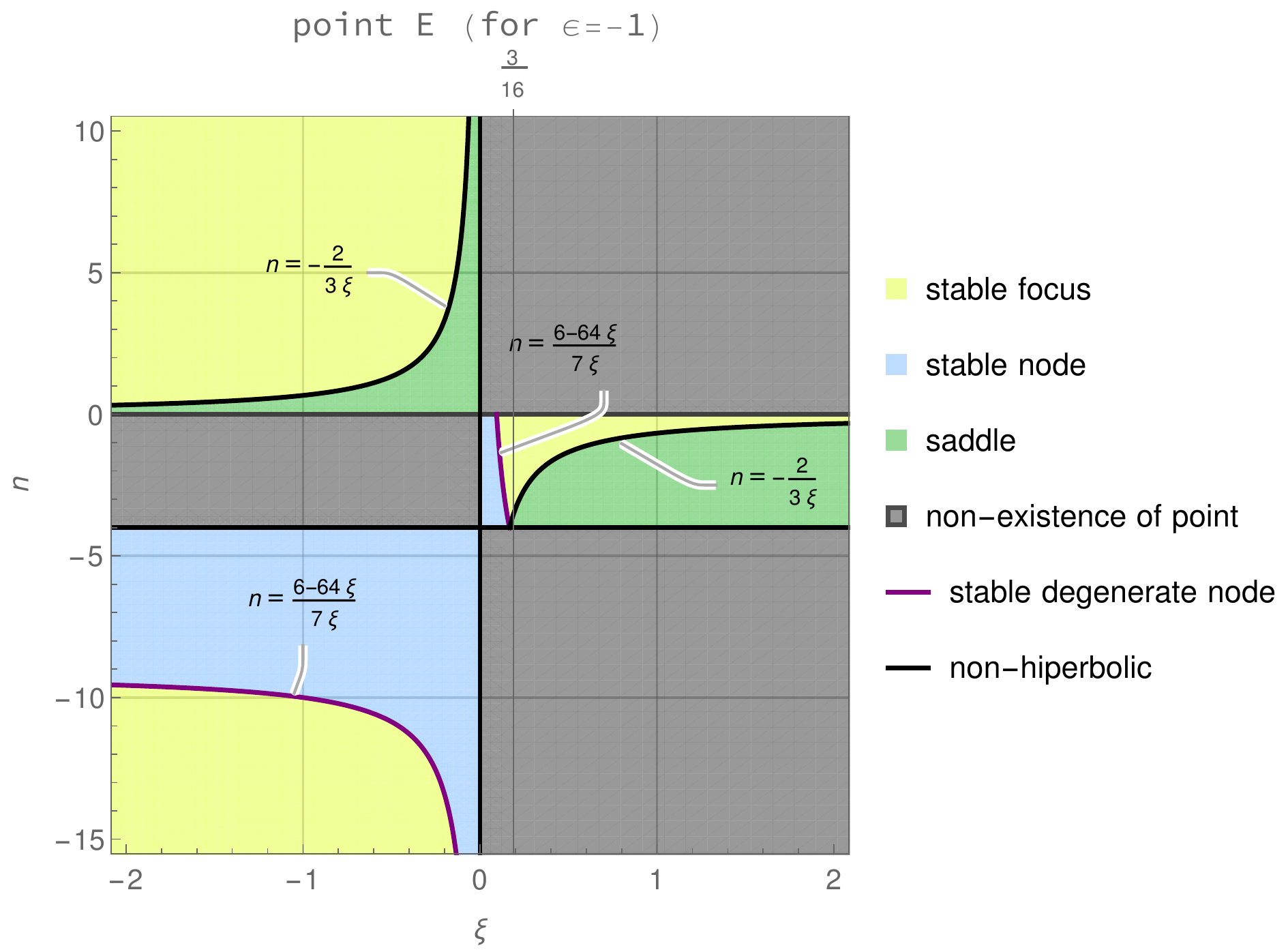}
	\vspace{-10px}
	\caption{Bifurcation diagram of the local stability of point $E$ for $\varepsilon=-1$.}
	\label{bif-ee}
\end{figure}

As the phase space $(u,v)$ is infinite, we will investigate firstly the finite region, and then go to the description at infinity. Finite equilibria of the system (\ref{system}) together with existence conditions (resulting from the fact that $u,v\in\mathbb R$) are shown in Table \ref{critical-points}, while eigenvalues and values of the equation of state parameter $w_\phi$ for equilibria are summarised in Table \ref{critical-points-details}. Figures \ref{bif-a}--\ref{bif-ee} present, in turn, bifurcation diagrams of the local stability of equilibrium points $A$, $B$, $C$ and $E$, whose stability properties depend on values of parameters $\xi$, $n$ and $\varepsilon$. Moreover, in these diagrams, there are put formulae for one-dimensional boundaries separating areas of the local topological equivalence.

We notice from Table \ref{critical-points} that, apart from the equilibrium line $F$, finite equilibrium points are located only on the axes of phase space. If for a finite point $v=0$ but $u\neq0$, then $\varepsilon\Omega_{\phi,\text{kin}}\rightarrow+\infty$. According to equation (\ref{energy_conservation_uv}), provided the condition (\ref{condition}) is satisfied, we have $\Omega_{\phi,\text{pot}}\rightarrow+\infty$. On the other hand, when $u=0$ and $v\neq0$, then $\Omega_{\phi,\text{kin}}=0$ and $\Omega_{\phi,\text{pot}}=1-6\varepsilon\xi v^{-2}$, which is greater than zero when the condition (\ref{condition}) is satisfied.

The region (\ref{condition}) has its boundaries at $v=0$ always at points $A$ and $B$. The boundary of this region is the limit of the physicality of the system. If $v\neq0$ we have $H^2\rightarrow+\infty$ and $|\dot\phi|\rightarrow+\infty$ on this boundary. If the line $F$ exists, the boundary has its extremum on this line at the point $u=-6\xi$. We denote this point as $K$.

In this paper, we will be investigating evolutionary scenarios of the universe, starting and finishing at the de~Sitter state. The {\it de~Sitter universe} is the solution to the Einstein field equations which assumes the dynamics of the universe to be dominated by the cosmological constant $\Lambda$, so the matter component (both baryonic and dark) is neglected. In this model, the pressure $p$ and the energy density $\rho$ satisfy
\begin{equation}
	p_{dS}=-\rho_{dS}=-\frac{\Lambda}{\kappa^2}=\mathrm{const},
\end{equation}
so the equation of state parameter $w$ is
\begin{equation}
	w_{dS}=-1.
\end{equation}
For the spatially flat universe ($k=0$) the scale factor $a$, within the de~Sitter model, depends on time as
\begin{equation}
	a(t)\propto e^{\pm\sqrt\frac{\Lambda}{3}t}.
\end{equation}
Solutions with the `$-$' sign in the exponent are frequently called {\it anti-de~Sitter states} in contrast to the solutions with the `$+$' sign, called {\it de~Sitter states}. The Hubble function $H$ in this case is equal
\begin{equation}
	H_{dS}=\pm\sqrt\frac{\Lambda}{3}=\mathrm{const}.
\end{equation}

By looking at the de~Sitter universe presence conditions in Table~\ref{critical-points-details} we can distinguish, using bifurcation diagrams (Figures~\ref{bif-a}--\ref{bif-ee}), sets of parameters for which an evolution from the initial (stable in the past) de~Sitter to the final (stable) de~Sitter state is possible. We can divide these sets into two groups: representing generic and non-generic de~Sitter--de~Sitter evolution. The generic evolution takes place, when there exists a family of solutions favouring given conditions, while the non-generic evolution takes place if only one particular solution corresponds to conditions. For example, when a family of orbits comes out of a stable in the past equilibrium (like an unstable node or a focus) representing the de~Sitter state ($w=-1$), and then finishes in a stable point also representing the de~Sitter state, it is generic evolution. On the other hand, if for example a starting or a final $w=-1$ point is a saddle, then there exists only one trajectory---a separatrix---which is able to reach this saddle (in the past or in the future, respectively), so the evolution is non-generic. Sets of parameters for which the evolution from the de~Sitter state to the de~Sitter state occurs have been shown in Table \ref{des_evolution_conditions}.

\newpage

\begin{table}[ht!]
	\captionsetup{font=small}
	\small
	\centering
	\caption{Sets of parameters for which the universe undergoes the evolution starting from the de~Sitter state and finishing in the de~Sitter state.}
	\label{des_evolution_conditions}
	\vspace{-10px}
	\begin{tabular}{@{}llllll@{}}
		\toprule
		no. & $\xi$ & $n$ & $\varepsilon$ & starting point & final point \\
		\midrule
		\multicolumn{6}{l}{1. Generic de~Sitter--de~Sitter evolution} \\
		\midrule
		(a) & $\frac{3}{16}$ & $(0,+\infty)$ & $+1$ & unstable node $A$ & stable focus $E$ \\
		(b) & $\frac{3}{16}$ & $(-2,0)$ & $-1$ & unstable node $A$ & stable focus $E$ \\
		(c) & $\left(\frac{3}{16},\frac{1}{4}\right)$ & $-2$ & $-1$ & unstable node $C$ & stable focus $E$ \\
		\midrule
		\multicolumn{6}{l}{2. Non-generic de~Sitter--de~Sitter evolution} \\
		\midrule
		(d) & $(-\infty,0]$ & $-2$ & $+1$ & saddle $E$ & stable node $C$ \\
		(e) & $\frac{3}{16}$ & $\left(-3\frac{5}{9},-2\right)$ & $-1$ & saddle $A$ & stable focus $E$ \\
		(f) & $\frac{3}{16}$ & $\left[-4,-3\frac{5}{9}\right)$ & $-1$ & saddle $A$ & saddle $E$ \\
		(g) & $\left[0,\frac{3}{16}\right)$ & $-2$ & $-1$ & saddle $C$ & stable node/focus $E$ \\
		(h) & $\left(\frac{1}{3},+\infty\right)$ & $-2$ & $-1$ & saddle $E$ & stable node $C$ \\
		\bottomrule
	\end{tabular}
\end{table}

\begin{figure}[t!]
	\captionsetup{font=small}
	\small
	\centering
	\includegraphics[width=\textwidth]{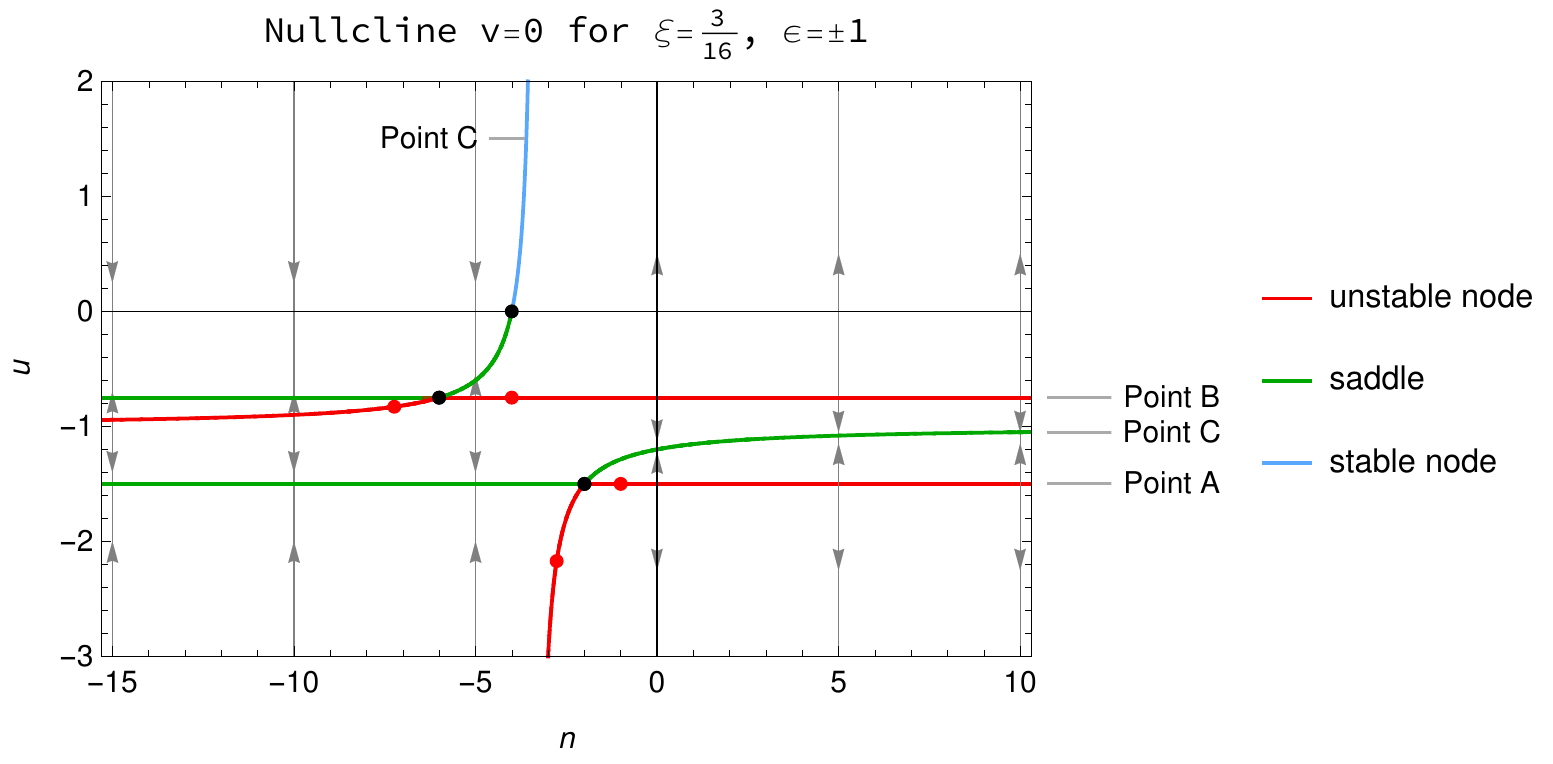}
	\caption{Bifurcation diagram of the local stability of equilibrium points of system (\ref{system}) on the $v=0$ nullcline under variation of the parameter $n$ for fixed $\xi=\frac{3}{16}$ and $\varepsilon=\pm1$. Black dots in this diagram denote a non-hyperbolic equilibrium point, while red dots denote an unstable proper node. Grey arrows show the direction of the flow on the nullcline.}
	\label{bif-v0}
	\vspace{15px}
	\includegraphics[width=\textwidth]{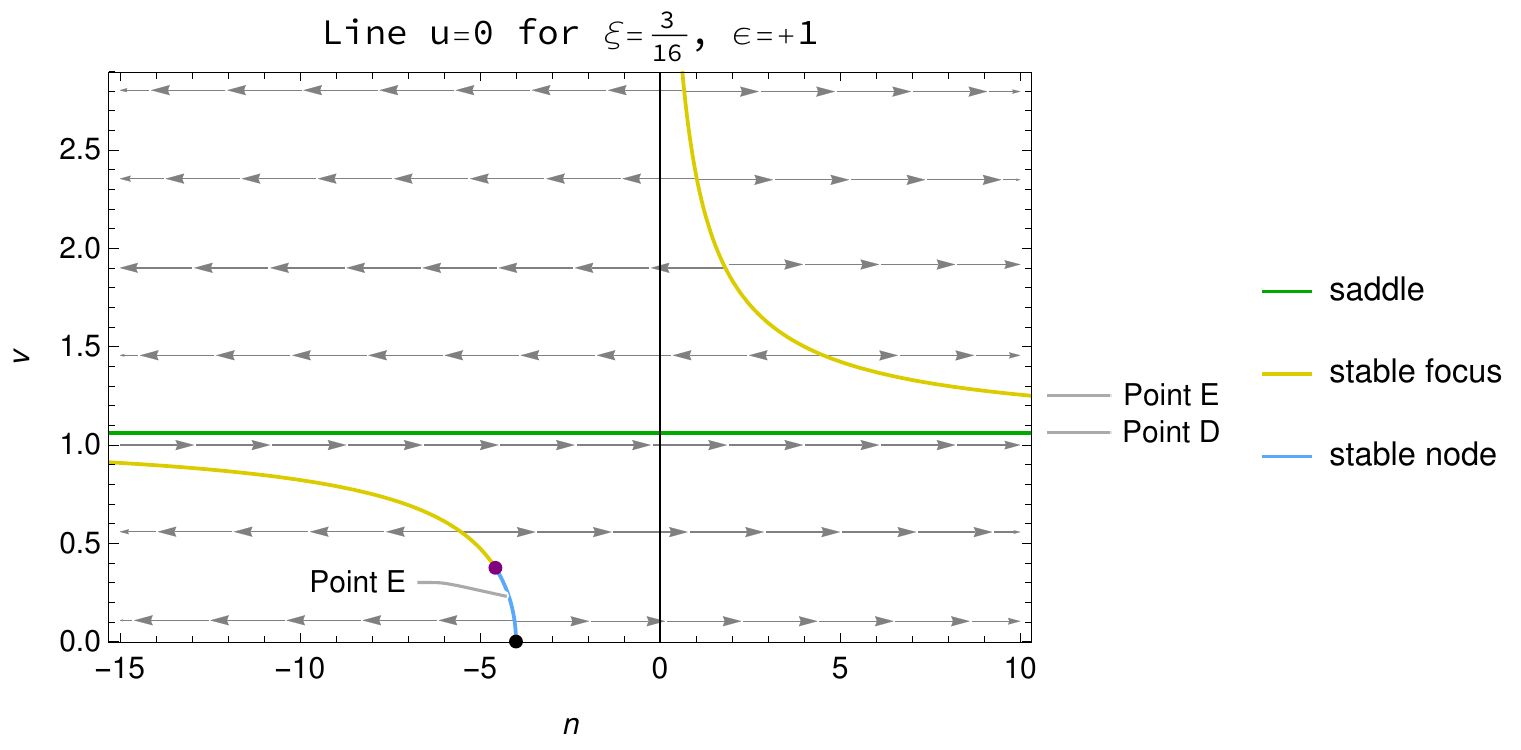}
	\caption{Bifurcation diagram of the local stability of equilibrium points of system (\ref{system}) on the $u=0$ line under variation of the parameter $n$ for fixed $\xi=\frac{3}{16}$ and $\varepsilon=+1$. The black dot in this diagram denotes a non-hyperbolic equilibrium point, while the purple dot denotes a stable degenerate node. Grey arrows show the direction of the flow on the $u=0$ line---arrows directed to the right denote the flow following towards increasing values of $u$, whereas arrows directed to the left denote the flow following towards decreasing values of $u$. The flow on the $u=0$ line is perpendicular to this line since, on it, $v'=0$ in system (\ref{system}).}
	\label{bif-u0}
\end{figure}

\begin{figure}[t!]
	\captionsetup{font=small}
	\small
	\centering
	\includegraphics[width=\textwidth]{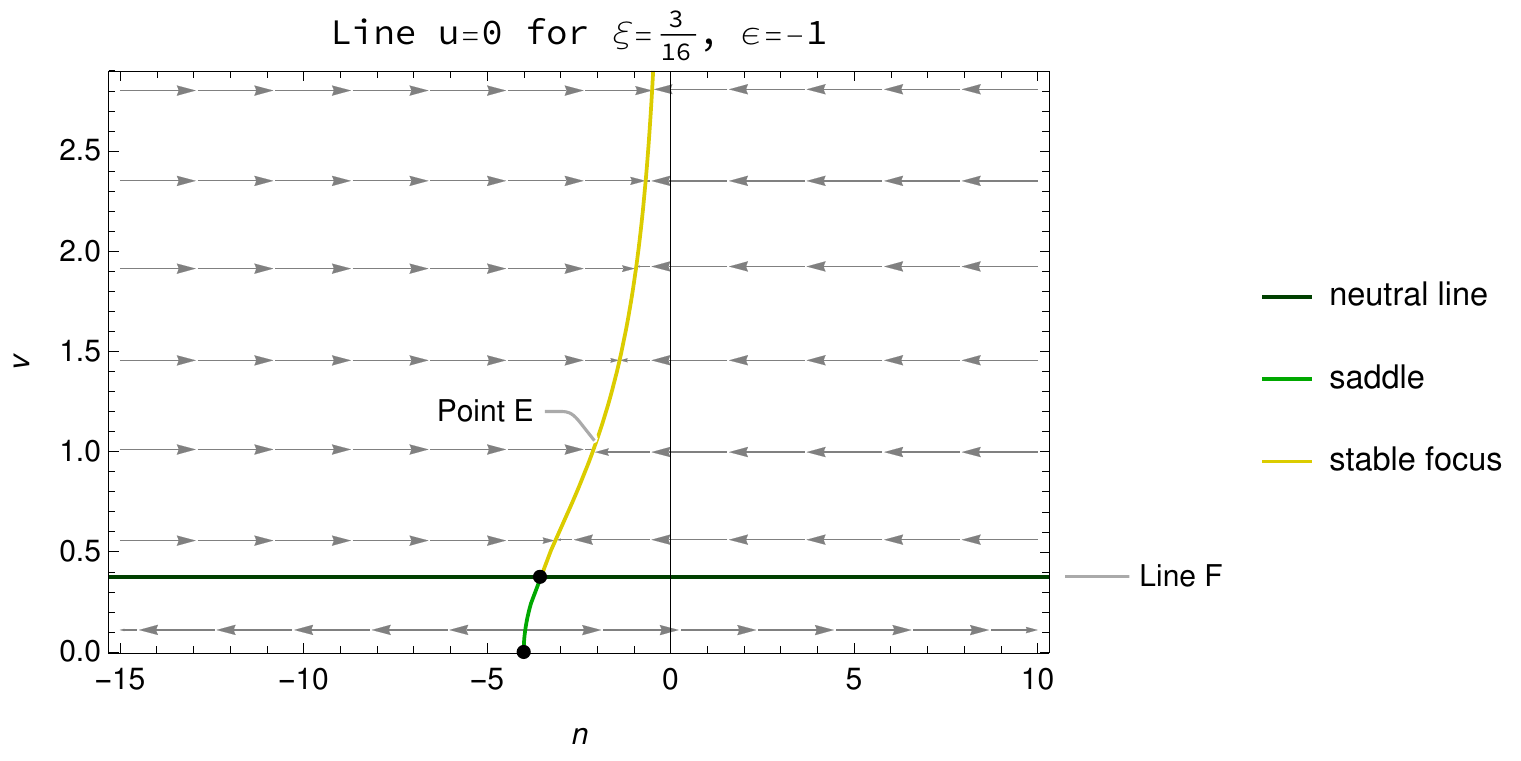}
	\caption{Bifurcation diagram of the local stability of equilibrium points of system (\ref{system}) on the $u=0$ line under variation of the parameter $n$ for fixed $\xi=\frac{3}{16}$ and $\varepsilon=-1$. Black dots in this diagram denote a non-hyperbolic equilibrium point. Grey arrows show the direction of the flow on the $u=0$ line---arrows directed to the right denote the flow following towards increasing values of $u$, whereas arrows directed to the left denote the flow following towards decreasing values of $u$. The flow on the $u=0$ line is perpendicular to this line since, on it, $v'=0$ in system (\ref{system}).}
	\label{bif-u0e}
\end{figure}

We will analyse generic evolutionary scenarios found in this work since they are more interesting than non-generic ones. Let us start with a discussion of bifurcations for these scenarios. Figures \ref{bif-v0}--\ref{bif-u0e} present bifurcation diagrams of the local stability of equilibria on the $v=0$ nullcline and $u=0$ line under variation of the parameter $n$ for fixed $\xi=\frac{3}{16}$ and $\varepsilon=\pm1$. This corresponds to generic evolutionary scenarios (a) and (b) in Table \ref{des_evolution_conditions}.

In Figure \ref{bif-v0} we see the occurrence of two transcritical bifurcations on the $v=0$ nullcline. The first takes place for $n=-6$ (belongs to the general bifurcation condition $n=-4-\sqrt{6\xi(6\xi-1)}/\xi$) and concerns points $B$ and $C$. For $n<-6$ the point $B$ is a saddle and the point $C$ is an unstable node, while after the `collision' they exchange their stability properties. The second transcritical bifurcation happens for $n=-2$ (belongs to the general condition $n=-4+\sqrt{6\xi(6\xi-1)}/\xi$) and concerns points $A$ and $C$. The situation looks similarly as in the previous case: for $n<-2$ the point $A$ is a saddle and the point $C$ is an unstable node, and these properties are exchanged between points after the bifurcation.

The local bifurcation diagram on the $u=0$ line for $\varepsilon=+1$ is shown in Figure~\ref{bif-u0}. The only transition visible in this diagram refers to the point $E$ which undergoes a change from a stable focus via a stable degenerate node (the purple dot) for $n=-4\frac{4}{7}$ (in general $n=(6-64\xi)/(7\xi)$) into a stable node. In fact, this transition is not a bifurcation, because---despite the fact that flows near nodes and foci are neither orbitally nor smoothly equivalent---they are topologically equivalent, which has been shown in Example 2.1 of \cite{kuznetsov98}.

Figure \ref{bif-u0e} presents the local bifurcation diagram on the $u=0$ line for $\varepsilon=-1$. For $n=-3\frac{5}{9}$ (in general $n=-2/(3\xi)$) another bifurcation appears when the point $E$ coalesces with the neutral line $F$ just to separate again as $n$ exceeds this bifurcation value. Point $E$ passes at this value of $n$ from a saddle into a stable focus, while the line $F$ still remains neutral, but the flow in its neighbourhood near $u=0$ reverses its direction---before the bifurcation the flow follows towards decreasing values of $u$ below the line $F$ and increasing values of $u$ above the line $F$, while after the bifurcation the flow, as well below as above the line $F$, reverses.

Furthermore, combining Figures \ref{bif-v0}, \ref{bif-u0} and \ref{bif-u0e}, we can notice that a bifurcation takes place also for $n=-4$ in the point $(u,v)=(0,0)$. For $\varepsilon=+1$, as $n$ approaches $-4$ from below, the saddle $C$ on the negative part of the $u=0$ nullcline and the stable node $E$ on the positive part of the $v=0$ axis are approaching the $(0,0)$ point. They both reach that point for $n=-4$, and then, when $n$ exceeds that value, the point $C$ becomes a stable node (intercepts the stability from the point $E$) and is moving towards increasing values of $u$, while the point $E$ disappears (its coordinates become complex). It is so to speak a combination of the saddle-node and the transcritical bifurcation: the point $E$ behaves as in the saddle-node bifurcation, while the point $C$ behaves as in the transcritical bifurcation. In turn, for $\varepsilon=-1$ the bifurcation is reversed. For $n<-4$ there is only the saddle $C$ on the negative part of $v=0$ axis, which reaches the point $(0,0)$ for $n=-4$, and when $n$ exceeds $-4$, it appears the saddle $E$, moving towards increasing values of $v$, while the point $C$ becomes a stable node (gives its instability to the point $E$) moving towards increasing values of $u$. In that case, as well as in the previous one, the point $E$ behaves as in the saddle-node bifurcation and the point $C$ behaves as in the transcritical bifurcation. In general, these transitions take place for all values of $\xi$ different than $0$ and $\frac{1}{6}$.

Let us remark that within the range of parameters of both the generic evolutionary scenarios (a) and (b) there does not occur any bifurcation, therefore it is sufficient to prepare only one phase portrait for each scenario, and this portrait is representative for the whole set of parameters of a given scenario. Also for the non-generic scenarios (e) and (f), we see no bifurcation.

\begin{figure}[t!]
	\captionsetup{font=small}
	\small
	\centering
	\includegraphics[width=\textwidth]{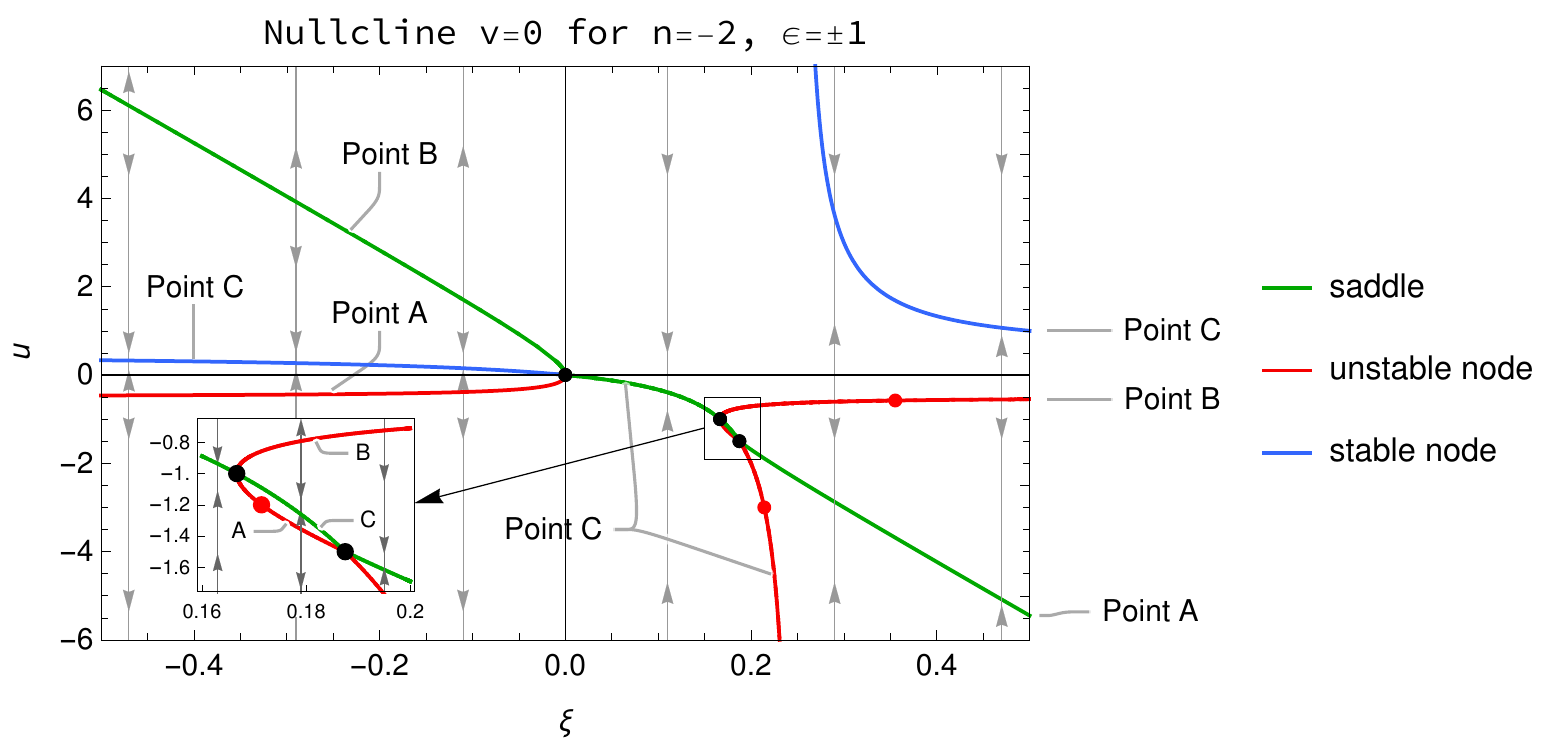}
	\caption{Bifurcation diagram of the local stability of equilibrium points of system (\ref{system}) on the $v=0$ nullcline under variation of the parameter $\xi$ for fixed $n=-2$ and $\varepsilon=\pm1$. Black dots in this diagram denote a non-hyperbolic equilibrium point, while red dots denote an unstable proper node. Grey arrows show the direction of the flow on the nullcline.}
	\label{bif-v0x}
	\vspace{20px}
	\includegraphics[width=\textwidth]{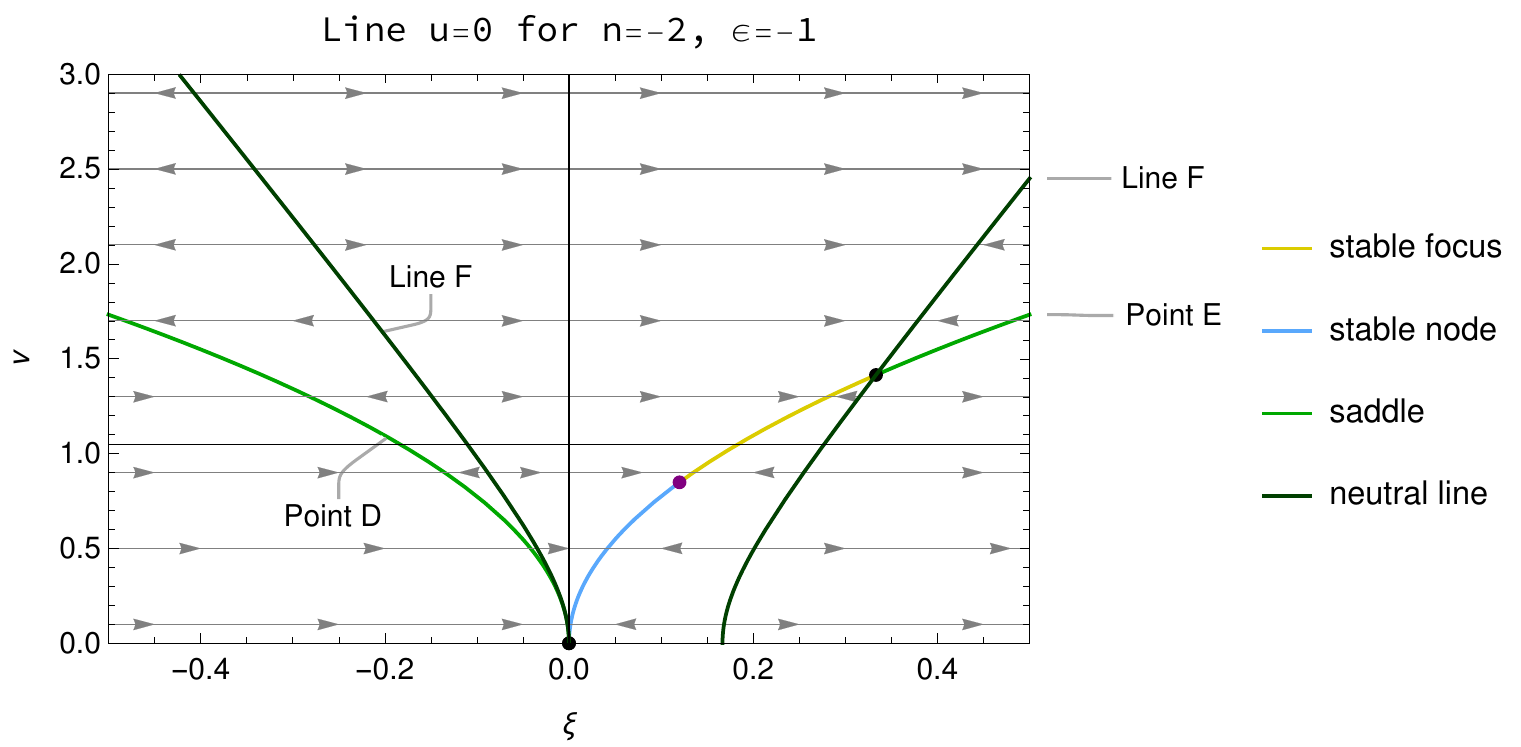}
	\caption{Bifurcation diagram of the local stability of equilibrium points of system (\ref{system}) on the $u=0$ line under variation of the parameter $\xi$ for fixed $n=-2$ and $\varepsilon=-1$. The black dot in this diagram denotes a non-hyperbolic equilibrium point, while the purple dot denotes a stable degenerate node. Grey arrows show the direction of the flow on the $u=0$ line---arrows directed to the right denote the flow following towards increasing values of $u$, whereas arrows directed to the left denote the flow following towards decreasing values of $u$. The flow on the $u=0$ line is perpendicular to this line since, on it, $v'=0$ in system (\ref{system}).}
	\label{bif-u0xe}
\end{figure}

Bifurcation diagrams on both $u$ and $v$ axes under variation of the parameter $\xi$ are presented in Figures \ref{bif-v0x}--\ref{bif-u0xe}. The choice of the values of fixed parameters, i.e., $n=-2$ and $\varepsilon=-1$, corresponds to generic evolutionary scenario (c). 

In Figure \ref{bif-v0x} we see the occurrence of two pitchfork bifurcations on the $v=0$ nullcline under variation of $\xi$. The first, taking place for $\xi=0$, is subcritical---stable point $C$ is surrounded by unstable points $A$ and $B$ for $\xi<0$, and at the critical value of $\xi$ they coalesce into one unstable (saddle) point $C$ (this bifurcation emerges in general at $\xi=0$ and $n>-4$). The second pitchfork bifurcation at $\xi=\frac{1}{6}$ appears when all points are unstable---two unstable nodes $A$ and $B$, surrounding the saddle $C$ emerge, as $\xi$ exceeds this critical value, from the single saddle $C$ (in general, this bifurcation takes place for $\xi=\frac{1}{6}$ and $n>-4$). Finally, for $\xi=\frac{3}{16}$ (belongs to the general bifurcation case $n=-4+\sqrt{6\xi(6\xi-1)}/\xi$), it occurs the transcritical bifurcation between points $A$ and $C$ from which, before the bifurcation, the first is an unstable node and the latter is a saddle. Bifurcation at this point can be seen also in Figure \ref{bif-v0}, however there the parameter $n$ which was varied.

On the line $u=0$, Figure \ref{bif-u0xe} shows other bifurcations under variation of $\xi$ for fixed $n=-2$ and $\varepsilon=-1$. The saddle $D$ and the neutral line $F$, which exist for $\xi<0$, coalesce in $\xi=0$. Above this value, $D$ and $F$ disappear and it appears the stable point $E$; the line $F$ emerges again at $\xi=\frac{1}{6}$. For $\xi=\frac{1}{3}$ the stable point $E$ and the neutral line $F$ coalesce and, as $\xi$ exceeds the bifurcation value, the point $E$ becomes a saddle, while the line $F$ remains neutral, but the flow below and above it reverses.

Combining Figures \ref{bif-v0x} and \ref{bif-u0xe} we see, that for $\xi=0$, $n=-2$ and $\varepsilon=-1$ all equilibrium points $A$--$F$ coalesce in $(u,v)=(0,0)$. It happens for any value of $n$ and $\varepsilon=\pm1$ at $\xi=0$ and refers to the case of the minimally coupled scalar field. Again, within the range of parameters of the generic scenario (c), we see no bifurcations, which means that one phase portrait will be representative for that case. In turn, the non-generic scenario (g) would require two portraits to be fully represented---the first for $0\leq\xi<\frac{1}{6}$ and the second for $\frac{1}{6}\leq\xi<\frac{3}{16}$, while in case of the scenario (h) one portrait would be enough.

\section{Phase portraits of the system}
\label{phase_portraits}

\begin{figure}[b!]
	\captionsetup{font=small}
	\small
	\centering
	\includegraphics[width=\textwidth]{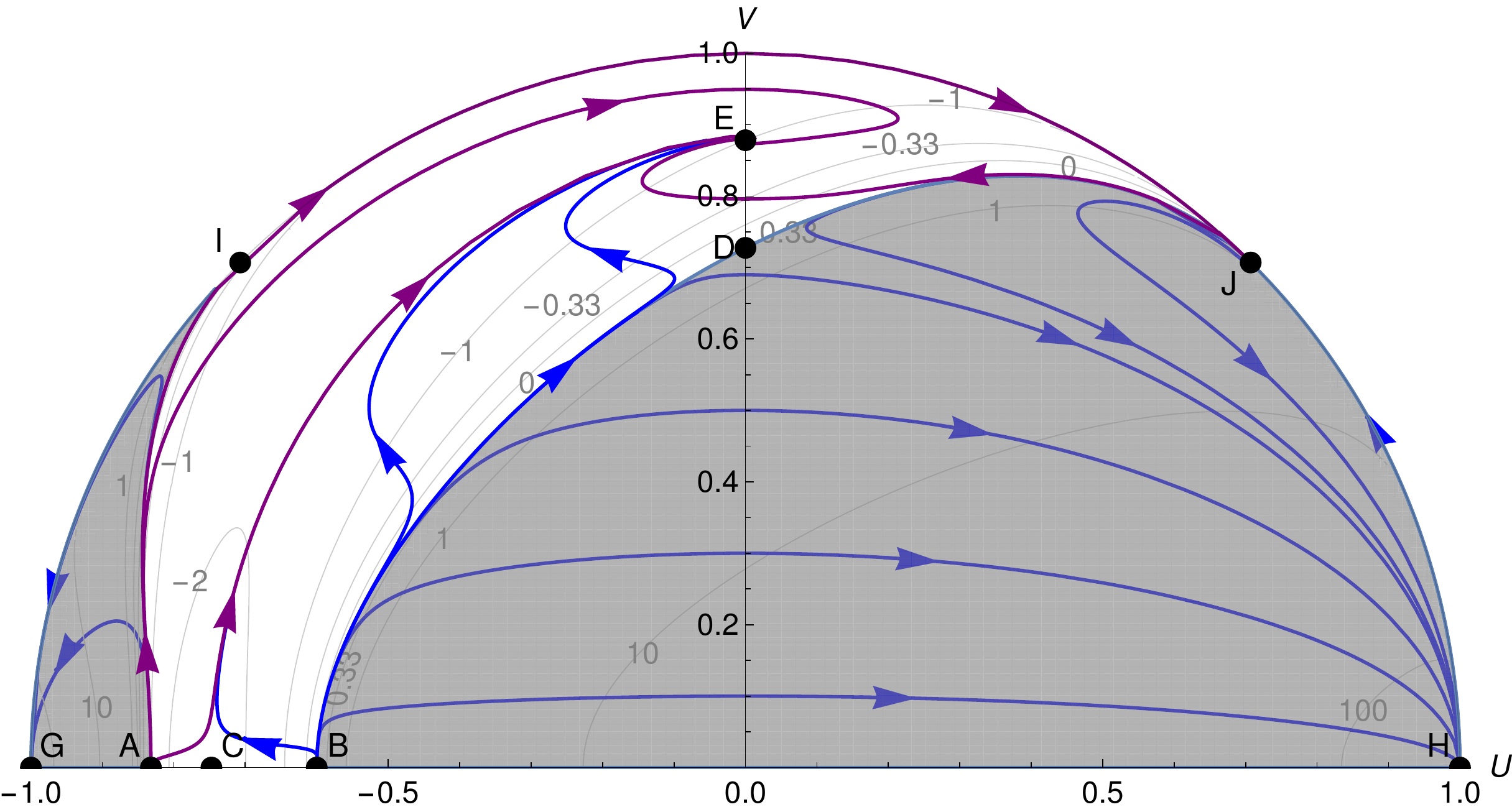}
	\caption{The phase portrait of system (\ref{system}) on the Poincar\'e sphere for the generic scenario (a). Values of parameters are $\xi=\frac{3}{16}$, $n=2$, $\varepsilon=+1$. Purple arrows denote orbits corresponding to the evolution from the de~Sitter state $A$ to the de~Sitter state $E$, while the blue arrows denote other orbits. Light grey contours correspond to values of the equation of state parameter $w_\phi$. The shaded region is non-physical.}
	\label{port-a}
\end{figure}

\begin{figure}[t!]
	\centering
	\includegraphics[width=\textwidth]{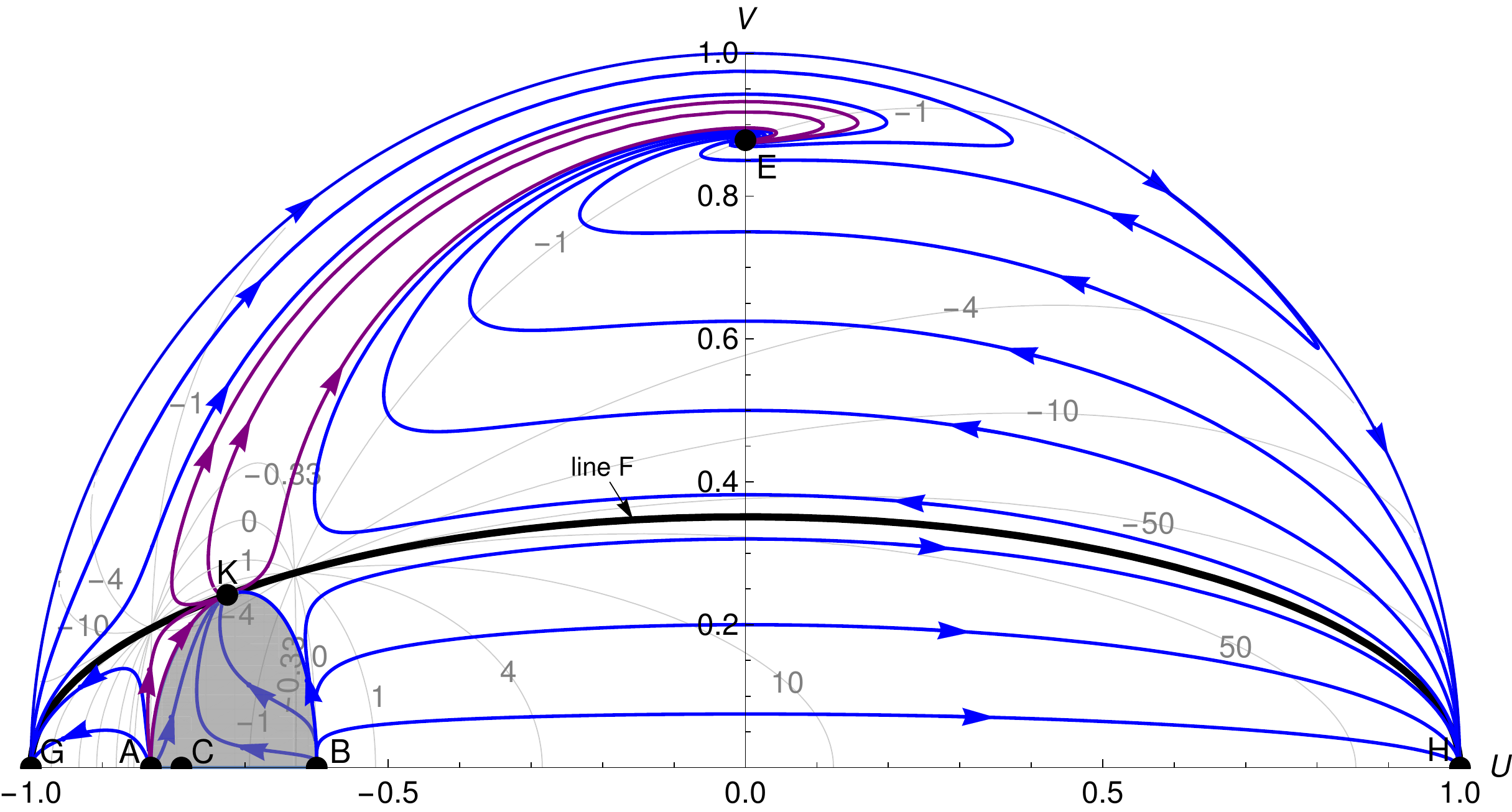}
	\caption{The phase portrait of system (\ref{system}) on the Poincar\'e sphere for the generic scenario (b). Values of parameters are $\xi=\frac{3}{16}$, $n=-1$, $\varepsilon=-1$. Purple arrows denote orbits corresponding to the evolution from the de~Sitter state $A$ to the de~Sitter state $E$, while the blue arrows denote other orbits. Light grey contours correspond to values of the equation of state parameter $w_\phi$. The shaded region is non-physical.}
	\label{port-b}
\end{figure}

\begin{figure}[ht!]
	\captionsetup{font=small}
	\small
	\centering
	\includegraphics[width=\textwidth]{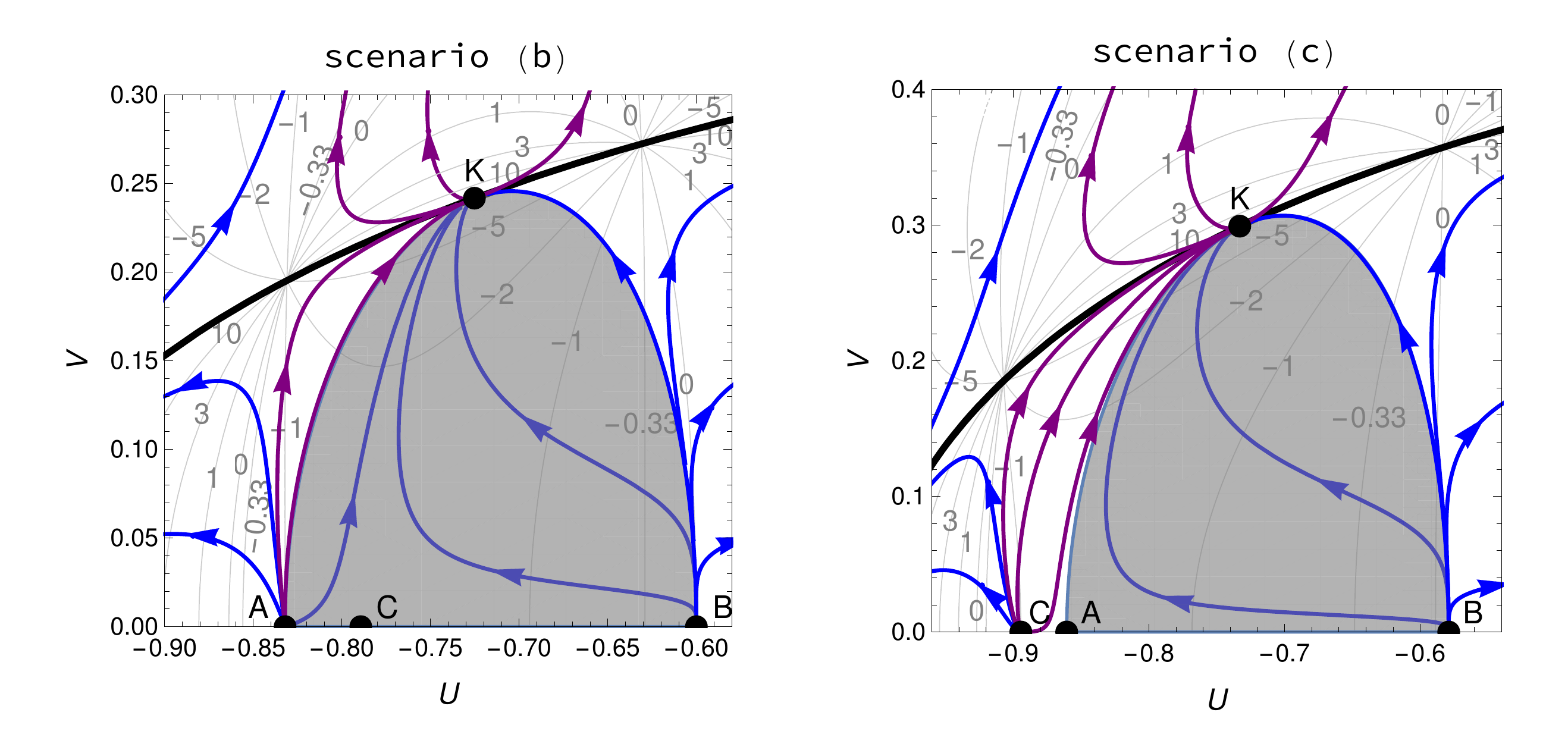}
	\vspace{-20px}
	\caption{The fragment of phase portraits of system (\ref{system}) on the Poincar\'e sphere for the generic scenario (b) (on the left) and the generic scenario (c) (on the right). Values of parameters are $\xi=\frac{3}{16}$, $n=-1$, $\varepsilon=-1$ for the (b) scenario, and $\xi=\frac{1}{5}$, $n=-2$, $\varepsilon=-1$ for the (c) scenario. Purple arrows denote orbits corresponding to the evolution from the de~Sitter state $A$ (left) or $C$ (right) to the de~Sitter state $E$, while the blue arrows denote other orbits. Light grey contours correspond to values of the equation of state parameter $w_\phi$. The shaded region is non-physical.}
	\label{port-c}
\end{figure}

Now, we are ready for a discussion of phase portraits of the system for generic de~Sitter--de~Sitter evolutionary scenarios. Figures \ref{port-a}--\ref{port-c} present phase portraits of system (\ref{system}) for the evolutionary scenarios (a)--(c). Each of them is representative for its case, as we noticed during the discussion of bifurcation diagrams of equilibria. Portraits have been drawn on the plane $(U,V)$, which is an orthographic projection of the northern hemisphere $W>0$ and the equator $W=0$ of the unit Poincar\'e sphere
\begin{equation}
	S^2 = \{(U,V,W) \in \mathbb R^3\ |\ U^2+V^2+W^2 = 1\},
\end{equation}
on which the $(u,v)$ infinite plane had been projected. According to the definition of the Poincar\'e sphere, coordinates $U$, $V$ are expressed by $u$, $v$ as follows \cite{perko91}
\begin{equation}
	U=\frac{u}{\sqrt{1+u^2+v^2}},\quad V=\frac{v}{\sqrt{1+u^2+v^2}},
\end{equation}
and the $U^2+V^2=1$ circle is corresponding to the infinity of the $(u,v)$ plane, i.e., points for which $\sqrt{u^2+v^2}\rightarrow+\infty$.

From an analysis of the flow of the system (\ref{system}) on the $(U,V)$ plane we obtain the information of the equilibria of the system at infinity; it has been summarised in Table~\ref{critical_points_infinity}.

\begin{table}[t!]
	\captionsetup{font=small}
	\small
	\centering
	\caption{Equilibrium points of the system (\ref{system}) at infinity together with existence conditions and values of the equation of state parameter.}
	\label{critical_points_infinity}
	\begin{tabular}{@{}ccccc@{}}
		\toprule
		point & $U$ & $V$ & existence & limit of $w_\phi$ \\
		\midrule
		$G$ & $-1$ & $0$ & \multirow{2}{*}{$\xi\in\mathbb R\ \land\ n\neq0\mbox*\ \land\ \varepsilon=\pm1$} & \multirow{2}{*}{$+\infty$ or** $-\infty$} \\
		$H$ & $1$ & $0$ & & \\
		$I$ & $-\frac{1}{\sqrt2}$ & $\frac{1}{\sqrt2}$ & \multirow{2}{*}{$\xi\in\mathbb R\ \land\ n\neq0\mbox*\ \land\ \varepsilon=+1$} & \multirow{2}{*}{$-4\xi+1$} \\
		$J$ & $\frac{1}{\sqrt2}$ & $\frac{1}{\sqrt2}$ & & \\
		\midrule
		\multicolumn{5}{p{0.7\textwidth}}{\scriptsize*For $n=0$ whole semicircle $V=\sqrt{1-U^2}$ is an equilibrium set.} \\
		\multicolumn{5}{p{0.7\textwidth}}{\scriptsize**The sign of the infinite limit of $w_\phi$ for $G$ and $H$ depends on values of $\xi$, $n$ and $\varepsilon$.} \\
		\bottomrule
	\end{tabular}
\end{table}

On the $V=0$ line we have, according to equations (\ref{energy_conservation_uv}) and (\ref{energy_parameters}), the kinetic energy parameter $\varepsilon\Omega_{\phi,\text{kin}}\rightarrow+\infty$. Moreover, at the point $G$, the density of energy related to the gravity-scalar field coupling has the limit $6\varepsilon\xi(1+2u)v^{-2}\rightarrow-\infty\cdot\sign(\varepsilon\xi)$, while at the point $H$ this limit is $+\infty\cdot\sign(\varepsilon\xi)$. On the $V=\sqrt{1-U^2}$ semicircle, if $V\neq0$, the kinetic energy has a finite value $\Omega_{\phi,\text{kin}}=\varepsilon U^2V^{-2}$, and the coupling part of the energy is equal to zero, thus if $\Omega_{\phi,\text{kin}}>1$, the point at infinity is non-physical, according to the condition (\ref{condition}). Hence, at points $I$, $J$ we have $\Omega_{\phi,\text{kin}}=\varepsilon$ and $\Omega_{\phi,\text{pot}}=1-\varepsilon$.

The phase portrait for the generic scenario (a), taking place for the canonical scalar field ($\varepsilon=+1$) with $\xi=\frac{3}{16}$ and $n>0$, is presented in Figure \ref{port-a}. Within the physical region~(\ref{condition}), the point $E$, representing the de~Sitter universe, is a global attractor to which trajectories from points $A$ ($w_\phi=-1$) and $B$ ($w_\phi=0$) are approaching. De~Sitter--de~Sitter trajectories (from $A$ to $E$---marked in purple) can be first either attracted by the saddle $C$ ($w_\phi<-\frac{9}{5}$) or both two saddles $I$ and $J$ in turn ($w_\phi=\frac{1}{4}$ both) before reaching the final point. As we can see, in this second case, the value $w_\phi=-\frac{1}{3}$ can be exceeded during evolution, which means the sign of $\ddot a$ can be changed to negative for a certain time.

The flow for the generic scenario (b), representing the phantom scalar field ($\varepsilon=-1$) for parameters values $\xi=\frac{3}{16}$ and $-2<n<0$, has been shown in Figure~\ref{port-b}, while in the left-hand side of Figure~\ref{port-c} enlargement of an area of the initial stage of an evolution has been outlined. In this case, an evolution from the de~Sitter to the de~Sitter state also starts at the point $A$ and finishes at the point $E$. These orbits are transiting through the line $F$ at the point $K$; when a trajectory is approaching this point from below (in the sense of values of the $V$ coordinate), the equation of state parameter $w_\phi\rightarrow-\infty$, and, after exceeding, $w_\phi$ decreases from the limit $+\infty$---at this point then, the system undergoes the so-called {\it $w$-singularity}. After passing the point $K$, orbits are heading towards the stable focus $E$.


Finally, the flow for the generic scenario (c), for which $\varepsilon=-1$, $\frac{3}{16}<\xi<\frac{1}{4}$, and $n=-2$, is topologically equivalent to the flow for scenario (b), as we can see from bifurcation diagrams---comparing Figure \ref{bif-u0e} with \ref{bif-u0xe} and Figure \ref{bif-v0} with \ref{bif-v0x} we notice that in the range of parameters for scenarios (b) and (c), mutual positions of equilibria are the same, except the pair of points $A$ and $C$, which undergoes a swap of positions and stability properties while passing between those two scenarios; we cannot, however, say it is a transcritical bifurcation of codimension 2 between these points, despite we have to vary values of both $\xi$ and $n$ parameters, because it is enough to change the value of one of these parameters to obtain the same bifurcation. Considering that the non-singular evolution starts from the point $A$ for scenario (b) and from the point $C$ for scenario (c), only the differences between portraits of the (b) and (c) cases are letters marking initial points of this kind of evolution and the shape of the non-physical region described by the condition \ref{condition} (it always has boundary at the point $A$). The last indicates that trajectories from the initial de~Sitter point to the point $K$ being attracted on their way by the saddle, which lies on the $V=0$ nullcline, are also physically acceptable in the case (c), in contrast to the case (b). The comparison of differences of phase portraits for the scenarios (b) and (c) has been presented in Figure~\ref{port-c}.

\section{Evolution of physical quantities}
\label{evolution_of_physical}

In this section, we will investigate the evolution of physical quantities, such as $H$, $\dot H+H^2$, $\Omega_{\phi,\text{kin}}$, $\Omega_{\phi,\text{pot}}$, over the cosmological time $t$ (the proper time of fundamental observers).

Let us notice firstly that the model we use allows us to determine explicitly only $H^2$, $\dot\phi^2$ and the quotient $\dot\phi/H$, which means that we are not able to calculate signs of $H$ and $\dot\phi$, because the model does not distinguish these signs\footnote{Notice that each of equations (\ref{friedmann}--\ref{klein_gordon}) contains $H^2$, $\dot\phi^2$ or the product $H\dot\phi$. Moreover, the new time, introduced in the system (\ref{system1}), is an increasing function of the cosmological time $t$ for $H>0$, and a decreasing function of $t$ for $H<0$, which prevents from recognition of signs of $H$ and $\dot\phi$ either directly from variables~$u$,$v$ or solutions of differential equations for $\dot H$ and $\ddot\phi$.}. Knowing that currently $H>0$, we can determine regions where $H$ goes to negative values by searching for $H^2=0$ points and checking a value of $\dot H+H^2$ there.

In order to present the time evolution of the scale factor $a$ and its derivatives, we need to transform equations (\ref{friedmann_uv}) and (\ref{acceleration_uv}); thus we obtain respectively
\begin{equation}
	H^2=M^{n+4}\left(\frac{\kappa}{\sqrt 6}\right)^{n+2}\frac{2v^{n+2}}{v^2-\varepsilon u^2-6\varepsilon\xi(1+2u)}
\end{equation}
and
\begin{equation}
	\frac{\ddot a}{a}=\dot H+H^2=-M^{n+4}\left(\frac{\kappa}{\sqrt 6}\right)^{n+2}\frac{v^{n+2}(3w_\phi+1)}{v^2-\varepsilon u^2-6\varepsilon\xi(1+2u)},
\end{equation}
where the overdot symbol ($\dot\ $) denotes a derivative with respect to $t$. The reparameterisation of time for dynamical system (\ref{system}) gives the new time $\tau$ which is related with $t$ by
\begin{equation}
	\mathrm dt=\mathrm d\tau\frac{\left[\frac{1}{3}\varepsilon v(\tau)^2-2\xi(1-6\xi)\right]^2}{H[u(\tau),v(\tau)]}.
\end{equation}
To reconstruct the time $t$ we need to determine the sign of $H$ and integrate the foregoing equation. In this work we assume arbitrarily $\tau=0$ such that $w_\phi(\tau\geq0)\approx-1$ at least to the third decimal place. Moreover, we put $t(\tau=0)=0$. Having the time~$t$, it is possible to calculate the scale factor directly from the definition of the Hubble function
\begin{equation}
	a(t)=\exp\left(\int\limits_0^tH(t')\mathrm dt'\right).
\end{equation}

\begin{figure}[t!]
	\captionsetup{font=small}
	\small
	\centering
	\includegraphics[width=\textwidth]{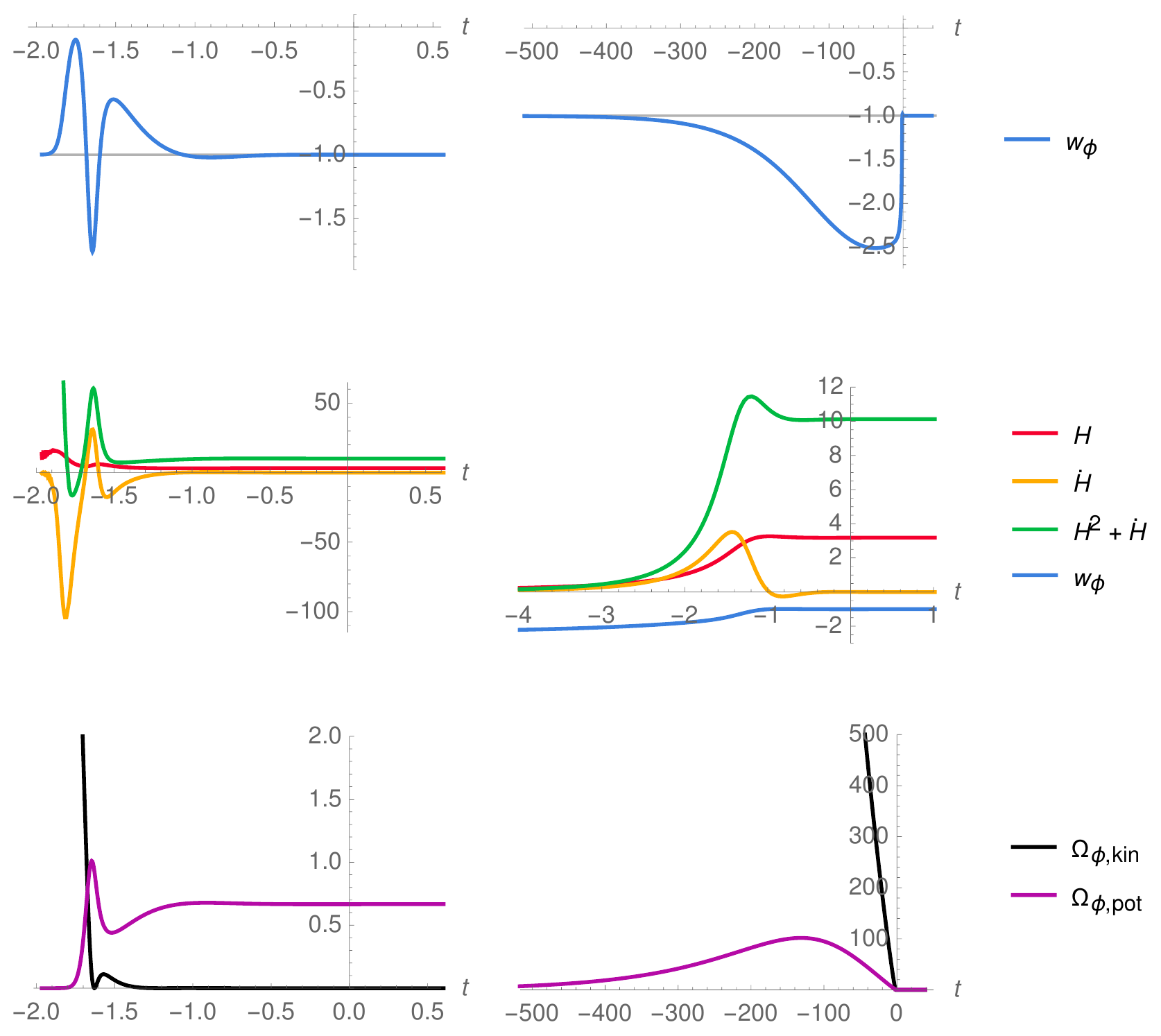}
	\caption{The evolution of physical quantities over the cosmological time $t$ for the evolutionary scenario (a) with values of parameters $\xi=\frac{3}{16}$, $n=2$, $\varepsilon=+1$. The left column corresponds to orbits passing near the equilibrium point $I$ (plots presented in the left column are for the orbit with the boundary conditions $U(\tau=-3.7)=-0.3$, $V(\tau=-3.7)=0.92$), while the right column corresponds to orbits passing near the equilibrium point $C$ (presented plots are for the orbit with boundary conditions $U(\tau=-910)=-0.822$, $V(\tau=-910)=0.012$). The time $t$ as well as quantities $H$, $H^2$ and $\dot H$ have been calculated in units, for which $M^{n+4}(\kappa/\sqrt6)^{n+2}=1$.}
	\label{physical_quantities-a}
\end{figure}

For every generic evolutionary scenario (a)--(c), there is no point on de~Sitter--de~Sitter orbits, for which $H^2=0$, with the exception of the initial point $A$ for (a) and (b) scenarios, where the limit of $H^2$ can be equal to zero\footnote{In the case (a) the limit of $H^2$ in the point $A$ is always equal to zero, while in the case (b) this limit is zero only if $-1<n<0$; for $-2<n<-1$ it diverges to infinity, while for $n=-1$ (then the point $A$ is the proper node) this limit depends on the direction of the approaching, nevertheless for physical de~Sitter--de~Sitter trajectories it is never equal to zero.}. This means, that all scenarios represent the expanding universe ($H>0$) at any finite time. Since $H=0$ corresponds to the static universe, scenarios (a) for any $n$ and (b) for $-1<n<0$ represent the initially static universe, which means that the scale factor $a>0$ at $t\rightarrow-\infty$. In case of scenarios (b) for $-2<n\leq-1$ and (c) for any $\xi$ it is $a\rightarrow0$ at $t\rightarrow-\infty$.

The evolution of physical quantities over the cosmological time $t$ for the evolutionary scenario (a) has been presented in Figure \ref{physical_quantities-a}. The left column corresponds to an orbit passing near the equilibrium point $I$, whereas the right column corresponds to an orbit passing near the equilibrium point $C$. In plots we see the occurrence of initial static state ($w_\phi=-1$, $\dot H=0$ and $H=0$---from the side of $t\rightarrow-\infty$), and final de~Sitter state ($w_\phi=-1$, $\dot H=0$ and $H=const>0$---from the side of $t\rightarrow+\infty$), between which a non-exponential evolution takes place. For trajectories attracted by the point $I$ it is possible for the universe to undergo a decelerated expansion ($w_\phi>-1/3$, $\dot H+H^2=\ddot a/a<0$) for a particular interval of time. The kinetic energy density $\Omega_{\phi,\text{kin}}$ is changing from $+\infty$ to $0$ during the evolution.

\begin{figure}[t!]
	\captionsetup{font=small}
	\small
	\centering
	\includegraphics[width=\textwidth]{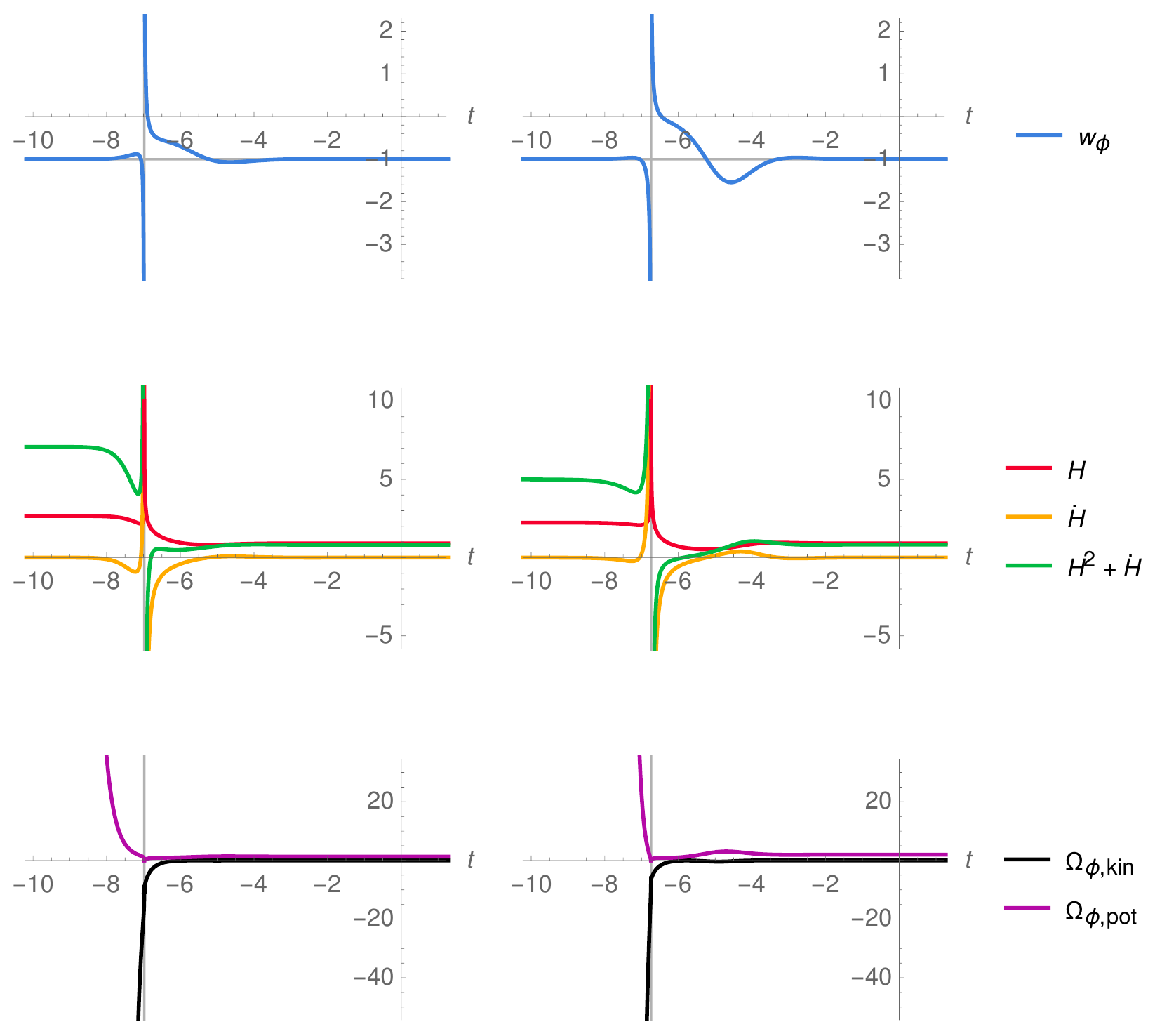}
	\caption{The evolution of physical quantities over the cosmological time $t$ for evolutionary scenarios (b) (in the left) and (c) (in the right). Values of parameters for plots in the left column are $\xi=\frac{3}{16}$, $n=-1$, $\varepsilon=-1$, while for plots in the right column are $\xi=\frac{1}{5}$, $n=-2$, $\varepsilon=-1$. Plots for the scenario (b) have been prepared for the orbit with the boundary conditions $U(\tau=-211)=-0.75$, $V(\tau=-211)=0.27$, whereas plots for the scenario (c) have been prepared for the orbit with the boundary conditions $U(\tau=-110)=-0.767$, $V(\tau=-110)=0.322$. The time $t$ as well as quantities $H$, $H^2$ and $\dot H$ have been calculated in units, for which $M^{n+4}(\kappa/\sqrt6)^{n+2}=1$. For the chosen boundary conditions, the intersection of the orbit with the equilibrium line $F$ (the point of the discontinuity of~$w_\phi$) takes place for $t=-6.97936$ in the case of scenario (b) or $t=-6.7425$ in the case of scenario~(c).}
	\label{physical_quantities-bc}
\end{figure}

Figure \ref{physical_quantities-bc} shows the evolution of physical quantities over the cosmological time $t$ for the evolutionary scenarios (b) (the left column) and (c) (the right column). Two de~Sitter states emerge from both sides of $t\rightarrow\pm\infty$ ($w_\phi=-1$, $\dot H=0$, $H=const>0$). During evolutions (b) and (c), it appears the discontinuity of $w_\phi$, which takes place when the orbit intersects the equilibrium line $F$. As the solution is approaching this line, the equation of state parameter $w_\phi\rightarrow-\infty$. After the solution exceeds line $F$, $w_\phi$ is decreasing from the limit $+\infty$; it appears the decelerated expansion ($H^2+\dot H<0$) for certain time. The kinetic energy $\Omega_{\phi,\text{kin}}$ is changing from $-\infty$ to $0$ throughout the evolution.

In Figures \ref{physical_quantities-a}--\ref{physical_quantities-bc}, we can see the presence of the initial exponential expansion, stretching to $t\rightarrow-\infty$, which models the inflation in the early universe. The second exponential expansion, stretching to $t\rightarrow+\infty$, describes the late time accelerated expansion, on the condition we add the matter content of the current universe. The full model, taking explicitly the matter component into account, can be an interesting aim of further research.

\section{Conclusions}
In the paper, we have used bifurcation theory methods to study the sensitivity of the model under a variation of the parameters. To show the power of these methods we have considered cosmological models which provide the explanation of the acceleration conundrum and are considered in the context of the inflation.

From our investigation, the following conclusions can be derived:
\begin{enumerate}
\item The application of bifurcation theory allow us to distinguish some class of emergent cosmological models from Einstein static universe and de Sitter universe. The first of these cases takes place for the canonical scalar field with the value of the non-minimal coupling parameter $\xi=\frac{3}{16}$ and the exponent appearing in the potential $n>0$, while two other cases concern the phantom scalar field for $\xi=\frac{3}{16}$ and $-2<n<0$ or for $\frac{3}{16}<\xi<\frac{1}{4}$ and $n=-2$. Furthermore, we have indicated five cases of non-generic de~Sitter--de~Sitter evolutionary scenarios for either canonical or phantom scalar field.

\item In the phase space, there are present evolutional paths without the initial singularity, allowing to explain both the inflation and the accelerated phase of the expansion at the late times.
\item There are two types of initial states from which the Universe starts the evolution. In the first scenario, the Universe is emergent from the de~Sitter state and in the second one, the Universe is emergent from the static universe.
\item From the bifurcation analysis, we have obtained a pair of the critical values of the parameters $(\xi, n)$ which corresponds to bifurcation values.
\item We have demonstrated that the methods of the bifurcation theory allows us to distinguish the regions of parameter values with the generic and non-generic evolution paths.
\end{enumerate}

\section*{Acknowledgement}
The authors are very grateful to Orest Hrycyna and Adam Krawiec for comments and suggestions.

\end{document}